\documentclass[twocolumn]{aastex62}
\usepackage{mathrsfs}
\usepackage{amsmath}
\usepackage{color}
\usepackage{comment}
\usepackage{url}

\received{December 6, 2018}
\revised{February 22, 2019}
\accepted{February 26, 2019}
\submitjournal{AAS (ApJ)}

\begin{document}
\title{Dust continuum emission and the upper limit fluxes of sub-millimeter water lines of the protoplanetary disk around HD 163296 observed by ALMA}
\author[0000-0003-2493-912X]{Shota Notsu}
\altaffiliation{Research Fellow of Japan Society for the Promotion of Science (DC1)}
\affiliation{Department of Astronomy, Graduate School of Science, Kyoto University, Kitashirakawa-Oiwake-cho, Sakyo-ku, Kyoto, Kyoto 606-8502, Japan; snotsu@kusastro.kyoto-u.ac.jp}

\author[0000-0002-5082-8880]{Eiji Akiyama}
\affiliation{Institute for the Advancement of Higher Education, Hokkaido University, Kita 17, Nishi 8, Kita-ku, Sapporo, Hokkaido 060-0817, Japan}

\author{Alice Booth}
\affiliation{School of Physics and Astronomy, University of Leeds, Leeds, LS2 9JT, UK}

\author[0000-0002-7058-7682]{Hideko Nomura}
\affiliation{Department of Earth and Planetary Science, Tokyo Institute of Technology, 2-12-1 Ookayama, Meguro-ku, Tokyo 152-8551, Japan}

\author[0000-0001-6078-786X]{Catherine Walsh}
\affiliation{School of Physics and Astronomy, University of Leeds, Leeds, LS2 9JT, UK}

\author[0000-0003-1659-095X]{Tomoya Hirota}
\affiliation{National Astronomical Observatory of Japan, 2-21-1 Osawa, Mitaka, Tokyo 181-8588, Japan}

\author[0000-0002-6172-9124]{Mitsuhiko Honda}
\affiliation{Department of Physics, School of Medicine, Kurume University, 67 Asahi-machi, Kurume, Fukuoka 830-0011, Japan}

\author{Takashi Tsukagoshi}
\affiliation{National Astronomical Observatory of Japan, 2-21-1 Osawa, Mitaka, Tokyo 181-8588, Japan}

\author[0000-0001-5178-3656]{T. J. Millar}
\affiliation{Astrophysics Research Centre, School of Mathematics and Physics, Queen's University Belfast, University Road, Belfast, BT7 1NN, UK}
\affiliation{Leiden Observatory, Leiden University, P.O. Box 9513, 2300 RA Leiden, The Netherlands}

\begin{abstract}
\noindent
In this paper, we analyse the upper limit fluxes of sub-millimeter ortho-H$_{2}$$^{16}$O 321 GHz, para-H$_{2}$$^{18}$O 322 GHz, and HDO 335 GHz lines from the protoplanetary disk around the Herbig Ae star HD 163296, using the Atacama Large Millimeter/Submillimeter Array (ALMA).
These water lines are considered to be the best candidate sub-millimeter lines to locate the position of the $\mathrm{H_2O}$ snowline, on the basis of our previous model calculations.
We compare the upper limit fluxes with the values calculated by our models with dust emission included,
and we constrain the line emitting region and the dust opacity from the observations.
We conclude that, 
if the outer edge of the region with high water vapor abundance and if the position of the water snowline are beyond 8 au also,
the mm dust opacity $\kappa_{\mathrm{mm}}$ will have a value larger than 2.0 cm$^{2}$ g$^{-1}$.
In addition, the position of the water snowline will be inside 20 au, if the mm dust opacity $\kappa_{\mathrm{mm}}$ is 2.0 cm$^{2}$ g$^{-1}$. 
Future observations of the dust continuum emission at higher angular resolution and sub-millimeter water lines with longer observation time are required to clarify the detailed structures and the position of the $\mathrm{H_2O}$ snowline in the disk midplane.
\end{abstract}


\section{Introduction}
\noindent 
\noindent Recently, high angular resolution and sensitivity observations of near-infrared dust scattered light (e.g., Gemini Planet Imager (GPI) on Gemini South and SPHERE on the Very Large Telescope) and sub-millimeter dust continuum emission (e.g., Atacama Large Millimeter/submillimeter Array (ALMA)) have found one or multiple gaps and rings for various protoplanetary disks.
The origins of these multiple gap and ring patterns are still under-discussion (see also Section 4), and the disk around HD 163296 is a great example of a disk that shows planet induced structures at multiple wavelengths.
\citet{Isella2016} observed the 232 GHz (1.3 mm, Band 6) dust continuum emission of the disk around HD 163296
with a spatial resolution of around 20 au
using ALMA, and revealed three dark concentric rings that indicate the presence of dust depleted gaps at about 50, 83, and 137 au from the central star (see also 
\citealt{Zhang2016, Andrews2018, Isella2018, Liu2018, Teague2018, Dent2019}).
\citet{Pinte2018} presented the detection of a large, localized deviation from Keplerian velocity in the $^{12}$CO J$= 2-1$ and J$= 3-2$ emission lines of this object obtained by ALMA, and the observed velocity pattern is consistent with the dynamical effect of a two-Jupiter-mass planet orbiting at a radius $\sim$260 au from the star.
In addition, \citet{Teague2018} found that the rotation curves of $^{12}$CO, $^{13}$CO, and C$^{18}$O J$= 2-1$ emission lines of this object obtained by ALMA had substantial deviations caused by local perturbations in the radial pressure gradient, which they explained as due to two Jupiter-mass planets at 83 and 137 au.
Several near-infrared dust scattered light observations \citep{Monnier2017, Guidi2018} for this object also detected ringed emission around 65 au, a position consistent with the first bright dust continuum ring observed by ALMA.
Here we note that the observation in the recent ALMA DSHARP of the 1.3mm dust continuum ($\Delta$$r\sim4$au) of the disk around HD 163296 found new small-scale structures, such as a dark gap at 10 au, a bright ring at 15 au, a dust crescent
at a radius of 55 au, and several fainter azimuthal asymmetries \citep{Andrews2018, Isella2018}.
\\ \\
Measuring the position of the water snowline (which corresponds to the sublimation front of water molecules, e.g., \citealt{Hayashi1981,Hayashi1985}) by observations in protoplanetary disks is important because it will constrain the dust-grain evolution and planet formation (e.g., \citealt{Oberg2011, Oka2011, Okuzumi2012, Ros2013, Banzatti2015, Piso2015, Piso2016, Cieza2016, Pinilla2017, Schoonenberg2017}), and the origin of water on terrestrial planets (e.g., \citealt{Morbidelli2000, Morbidelli2012, Morbidelli2016, Walsh2011, IdaGuillot2016, Sato2016, Raymond2017}).
Water ice features in disks have been detected through low dispersion spectroscopic observations \citep{Malfait1999, Terada2007, Terada2017, Honda2009, Honda2016, McClure2012, McClure2015, Min2016}.
However, it is difficult to directly locate the H$_{2}$O snowline through such water ice observations,
because the spatial resolution of existing telescopes is insufficient.
\\ \\
H$_{2}$O lines from disks have been detected through recent space infrared spectroscopic observations, such as {\em Spitzer}/IRS and {\em Herschel}/PACS, HIFI (for more details, see e.g., \citealt{Carr2008, Pontoppidan2010a, Hogerheijde2011, Salyk2011, Fedele2012, Fedele2013, Meeus2012, Riviere-Marichalar2012, Kamp2013, Podio2013, Zhang2013, vanDishoeck2014, Antonellini2015, Antonellini2016, Antonellini2017, Blevins2016, Banzatti2017, Du2017}). 
However, these lines mainly trace the disk surface and the cold water reservoir outside the H$_{2}$O snowline.
Water line profiles were detected by ground-based near- and mid-infrared spectroscopic observations using the Keck, VLT, and Gemini North/TEXES for some bright T Tauri disks \footnote[1]{In the remainder of this paper, we define the protoplanetary disks around T Tauri/Herbig Ae stars as ``T Tauri/Herbig Ae disks".} (e.g., \citealt{Salyk2008, Salyk2019, Pontoppidan2010b, Mandell2012}).
Those observations suggested that the hot water vapor resides in the inner part of the disks; however, the spatial and spectral resolution was not sufficient to investigate detailed structures, such as the position of the H$_{2}$O snowline.
In addition, the observed lines, with large Einstein A coefficients, are sensitive to the water vapor in the disk surface and are potentially polluted by slow disk winds.
\\ \\
In our previous papers \citep{Notsu2015, Notsu2016, Notsu2017, Notsu2018a}, on the basis of our calculations of disk chemical structures and water line profiles, we proposed how to identify the position of the H$_{2}$O snowline directly by analyzing the Keplerian profiles of water lines which can be obtained by high-dispersion spectroscopic observations
across a wide range of wavelengths (from mid-infrared to sub-millimeter, e.g., ALMA, SPICA).
We selected candidate water lines to locate the H$_{2}$O snowline based on specific criteria.
We concluded that lines which have small Einstein $A$ coefficients (A$_{\mathrm{ul}}$=$10^{-6} \sim10^{-3}$ s$^{-1}$) and relatively high upper state energies (E$_{\mathrm{up}}$$\sim$ 1000K) trace the hot water reservoir within the $\mathrm{H_2O}$ snowline, and can locate the position of the H$_{2}$O snowline.
In these candidate lines, the contribution of the optically thick hot midplane inside the H$_{2}$O snowline is large compared with that of the outer optically thin surface layer. 
This is because the intensities of lines from the optically thin region are proportional to the Einstein $A$ coefficient. Moreover, the contribution of the cold water reservoir outside the H$_{2}$O snowline is also small, because lines with high excitation energies are not emitted from the regions at a low temperature.
In addition, since the number densities of the ortho- and para-H$_{2}$$^{18}$O molecules are about 1/560 times smaller than their $^{16}$O analogues, they trace deeper into the disk than the ortho-H$_{2}$$^{16}$O lines (down to $z=0$), and lines with relatively smaller upper state energies ($\sim$ a few 100K) can also locate the position of the $\mathrm{H_2O}$ snowline.
Thus these H$_{2}$$^{18}$O lines are potentially better probes of the position of the H$_{2}$O snowline at the disk midplane, depending on the dust optical depth \citep{Notsu2018a}. 
\\ \\
The position of the H$_{2}$O snowline of a Herbig Ae disk exists at a larger radius compared with that around less massive and cooler T Tauri stars.
In addition, the position of H$_{2}$O snowline migrates closer to the star as the disk becomes older and mass accretion rate to the central star becomes smaller (e.g., \citealt{Oka2011, Harsono2015}).
Therefore, it is expected to be easier to observe the candidate water lines, and thus identify the location of the H$_{2}$O snowline, in Herbig Ae disks, younger T Tauri disks (e.g., HL Tau, \citealt{ALMA2015, Banzatti2015, Zhang2015, Okuzumi2016}), and disks around FU Orionis-type stars (e.g., V883 Ori, \citealt{Cieza2016, vantHoff2018}).
\\ \\
In this paper, we report our ALMA observations
of sub-millimeter water lines (ortho-H$_{2}$$^{16}$O 321 GHz, para-H$_{2}$$^{18}$O 322 GHz, and HDO 335 GHz) from the protoplanetary disk around Herbig Ae star HD 163296. 
These lines are considered to be the prime candidate water lines available at sub-millimeter wavelenghts to locate the position of the $\mathrm{H_2O}$ snowline \citep{Notsu2015, Notsu2016, Notsu2017, Notsu2018a}.
We also report the dust continuum emission from the disk around HD 163296 at a spatial resolution of around 15 au, that confirms the multi-ringed and gapped structure originally found in previous observations (e.g., \citealt{Isella2016, Dent2019}).
Section 2 outlines the observational setup, our data reduction process, and introduces the previous work on our target. 
The results and discussion about water line observations are described in Section 3, and those about dust continuum emission are reported in Section 4.
In Section 5, the conclusions are given.
\section{Observations}
\subsection{Observational setup and data reduction}
\noindent 
The high spatial-resolution continuum and water line observations at Band 7 ($\lambda\sim$0.9 mm) with ALMA was carried out in Cycle 3 on 2016 September 16 (2015.1.01259.S, PI: S. Notsu).
In the observation period, 40 of the 12m antennas were operational and the maximum baseline length was 2483.5m.
The correlater was configured to detect dual polarizations in four spectral windows with a bandwidth of 1.875 GHz and a resolution of 1953.125 kHz each,
resulting in a total bandwidth of 7.5 GHz.
The spectral windows were centered at 320.844 GHz (SPW1), 322.740 GHz (SPW2),
332.844 GHz (SPW3), and 334.740 GHz (SPW4), covering ortho-H$_{2}$$^{16}$O 321 GHz, para-H$_{2}$$^{18}$O 322 GHz, and HDO 335 GHz lines (see also Table 1).
The first local oscillator frequency (LO1) was tuned at 327.72597 GHz in order to avoid a strong atmospheric absorption around 325 GHz.
The integration time on source (HD 163296) is 0.723 hours (43.35 min.).
This integration time is around 20 \% of our requested time in our Cycle 3 proposal for the clear detection of water lines.
\\ \\
The phase was calibrated by observations of J1742-1517 and J1751-1950 approximately every 10 minutes, and J1733-1304 was used for absolute flux calibration. 
The observed passbands were calibrated by J1924-2914.
The visibility data were reduced and calibrated using the Common Astronomical Software Application (CASA) package, version 4.7.2.
The corrected visibilities were imaged using the CLEAN algorithm with briggs weighting with a robust parameter of 0.5 after
calibration of the bandpass, gain in amplitude and phase, and absolute flux scaling, and then flagging for aberrant data. 
The $uv$ sampling of our Band 7 data is relatively sparse. 
Since in this observation we focused on the water emission lines from the hot region inside the H$_{2}$O snowline ($\lesssim14$ au), we do not have good $uv$ coverage at shorter baselines.
From previous observations, it is known that the dust emission extends to $>$100 au.
Thus, in order to recover the missing flux, especially in the outer disk, we have combined Band 7 archival ACA (Atacama Compact Array) data (2016.1.00884.S, PI: V. Guzman) with our Band 7 data in our dust continuum imaging after applying a phase shift to account for proper motion and different input phase centers.
In addition to the usual CLEAN imaging, we performed self-calibration of the continuum emission to improve the sensitivity and image quality. The obtained solution table of
the self-calibration for the continuum emission was applied to the visibilities of the lines. 
The spatial resolutions of the final image is 0.174 $\times$ 0.124 arcsec with a
position angle (PA) of 76.799 $\deg$ for Band 7, corresponding to 17.7 au $\times$ 12.6 au.
The noise level (rms) of the Band 7 image is around 0.1 mJy beam$^{-1}$.
\\ \\
We also used ALMA Band 6 calibrated archival data (2013.1.00601.S, PI: A. Isella) of this object which was obtained in ALMA Cycle 2 to 
compare the dust continuum image between Bands 6 and 7.
We generate the Band 6 image with briggs weighting at a robust parameter of 0.5.
The spatial resolution of the final image is 0.278 $\times$ 0.184 arcsec with a
position angle (PA) of -87.987 $\deg$ for Band 6, corresponding to 28.2 au $\times$ 18.7 au.
The spatial resolution of this calibrated archival data is $\sim$1.7 times larger than that reported in \citet{Isella2016} (0.22 $\times$ 0.14 arcsec, corresponding to 22.3 au $\times$ 14.2 au), in which they generate the self-calibrated continuum image with different briggs
weighting parameter of -1 (uniform weighting).
%
The noise level (rms) of the calibrated archival data of Band 6 image is around 0.5 mJy beam$^{-1}$.
\\ \\
The total disk fluxes are 1.85Jy in Band 7, and 0.68Jy in Band 6, giving spectral index $\alpha_{\mathrm{mm}}$ of $\sim2.7$. 
This is in reasonable agreement with that measured by \citet{Pinilla2014} ($2.73\pm0.44$) and \citet{Guidi2016}, and a bit larger than that measured by \citet{Dent2019} ($2.1\pm0.3$).
\subsection{Target}
\noindent Our target HD 163296 is an isolated, young ($\sim$5 Myr), and intermediate mass ($\sim$2.3$M_{\bigodot}$) Herbig Ae star and has no evidence of a stellar binary companion. 
It is relatively nearby, and it is surrounded by a well-studied gas-rich disk with no hint of an inner hole (group II, e.g., \citealt{Honda2015}).
Here we note that according to the recent Gaia data release 2 \footnote[2]{https://www.cosmos.esa.int/web/gaia/dr2}, 
the distance obtained by Hipparcos measurement in the past ($d\sim$122 pc, e.g., \citealt{Perryman1997, vandenAncker1997}) was corrected to $d\sim$101.5 pc \citep{Gaia2018}.
In this paper, we adopt this new distance value (101.5 pc).
It has a relatively large inclination angle ($i\sim$42 deg, \citealt{Isella2016}), and thus we expected the characteristic double-peaked velocity profiles of gas in Keplerian rotation with large velocity widths ($\sim$30 km s$^{-1}$) due to the compact emitting area, which was suitable to detect the position of the H$_{2}$O snowline.
In addition, this object has been observed in many transitions at various wavelengths.
The spectrally resolved CO lines in the sub-millimeter show the characteristic double-peaked profiles of gas in Keplerian rotation (e.g., \citealt{Dent2005, Akiyama2011}). 
\\ \\
The CO snowline of this object is resolved directly using C$^{18}$O, N$_{2}$H$^{+}$, and DCO$^{+}$ line data obtained by ALMA 
(e.g.,  \citealt{Qi2011, Qi2015, Mathews2013, Salinas2017, Salinas2018}).
Previous ALMA observations showed that the continuum emission has a local maximum near the location of the CO snowline
\citep{Guidi2016, Zhang2016}.
The CO snowline position is around 75 au, using the new Gaia data.
In addition, the measurements of the spectral index $\alpha$ indicated the presence of large grains and pebbles ($\sim$ 1 cm) in the inner regions
of the disk (inside 40 au) and smaller grains, consistent with ISM sizes, in the outer disk (beyond 125 au), which would suggest a grain size distribution consistent with an enhanced production of large grains at the CO snowline and consequent transport to the inner regions \citep{Guidi2016}.
\citet{Boneberg2016} suggested by combining their C$^{18}$O line models, previous CO snowline observations, and spectral energy distributions, that the gas to dust mass ratio $g/d$ would be low ($<$ 20) within the CO snowline, the disk gas mass is $\sim(8-30)\times10^{-3}M_{\bigodot}$, and the mm dust opacity $\kappa_{\mathrm{mm}}$ is $\sim3$ cm$^{2}$ g$^{-1}$, assuming the values of C$^{18}$O abundances reported in star-forming clouds ($\sim10^{-7}-10^{-6}$). They also suggested that the value of $g/d$ would be larger (up to ISM-like value, $\sim$100) if they assume lower C$^{18}$O abundance because of CO depletion.
\\ \\
According to \citet{Reboussin2015} and \citet{Bosman2018}, the C atoms generated through CO photodissociation in the upper layers can be effectively removed through formation of species other than CO (e.g., CO$_{2}$ and CH$_{4}$). Photodissociation is normally localized in the disk surface, and the C$^{18}$O abundance may be affected only if the CO dissociating photons penetrate to the disk midplane, or if the surface continued to be depleted of CO over very long time-scales. 
In addition, significant depletion of CO will occur in the outer cold parts of the disks with a high cosmic-ray rate \citep{Bosman2018, Schwarz2018}.
Here we note that in one relatively old T Tauri disk around TW Hya, the abundances of CO and its isotopologues are observed as about 100 times lower than their ISM values \citep{Favre2013, Schwarz2016}.
\\ \\
\citet{Carney2019} derived the 3$\sigma$ disk-integrated intensity upper limits of methanol (CH$_{3}$OH) emission lines in ALMA Bands 6 and 7 toward the disk around HD 163296, and found that the disk is less abundant in methanol with respect to formaldehyde (H$_{2}$CO) compared to the disk around TW Hya.
They discussed possible reasons for the lower CH$_{3}$OH/H$_{2}$CO ratio, such as differences in the disk structure and/or CH$_{3}$OH and H$_{2}$CO desorption processes, uncertainties in the grain surface formation efficiency, and a higher gas-phase formation of H$_{2}$CO.
They estimated additional observation times required for ALMA detections of CH$_{3}$OH lines in the disk around HD 163296, depending on the different CH$_{3}$OH/H$_{2}$CO ratios.
\\ \\
The position of the H$_{2}$O snowline of a disk around a Herbig Ae star
with stellar luminosity of 36$L_{\bigodot}$
is $\sim$14 au from the central star, on the basis of our calculations \citep{Notsu2017, Notsu2018a}. 
Previous infrared observations of HD 163296 detected far-infrared water lines.
\citet{Meeus2012} and \citet{Fedele2012, Fedele2013} reported that three far-infrared ortho-H$_{2}$$^{16}$O emission lines (63.32, 71.95, 78.74 $\mu$m) are detected at slightly above 3$\sigma$ with Herschel/PACS.
These lines have large Einstein A coefficients (A$_{\mathrm{ul}}$$\sim$ 1 s$^{-1}$) and the spectral resolving power of the data was not high (R$\sim$$\lambda/\Delta\lambda\sim$1000-3000). 
They argued that
these H$_{2}$O lines are emitted from an upper hot water layer at radial distances $\sim$20 au, where water formation is driven by high-temperature neutral-neutral reactions. This argument is consistent with the results of our model calculations of lines with large Einstein A coefficients \citep{Notsu2016, Notsu2017, Notsu2018a}.
\begin{figure*}[htbp]
\begin{center}
\includegraphics[scale=0.65]{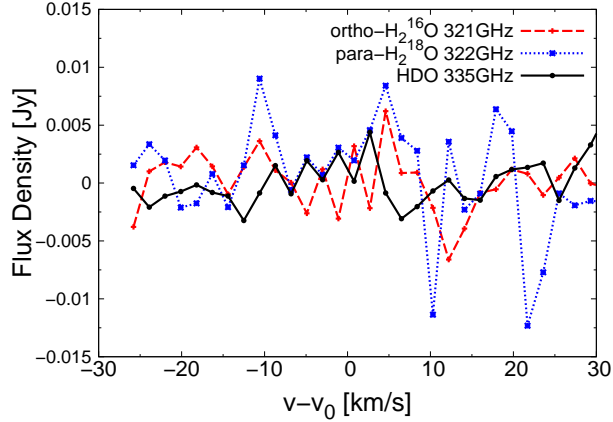}
\end{center}
\vspace{0.5cm}
\caption{
\noindent
The observed flux densities around the line centers of the ortho-$\mathrm{H_2}$$^{16}\mathrm{O}$ line at 321 GHz ({\it red dashed line with cross symbols}), the para-$\mathrm{H_2}$$^{18}\mathrm{O}$ line ({\it blue dotted line with filled square and cross symbols}) at 322 GHz, and the HDO line at 335 GHz ({\it black solid line with circle symbols}) of the disk around HD 163296, with a velocity resolution d$v=$1.9 km s$^{-1}$. In obtaining the observed line flux densities, we adopted a circular aperture with radius 20 au.
\vspace{0.3cm}
}\label{Figure1_paperIV}
\end{figure*} 
\section{Water line emission}
\subsection{Upper limit of the water line fluxes}
\begin{figure*}[htbp]
\begin{center}
\includegraphics[scale=0.6]{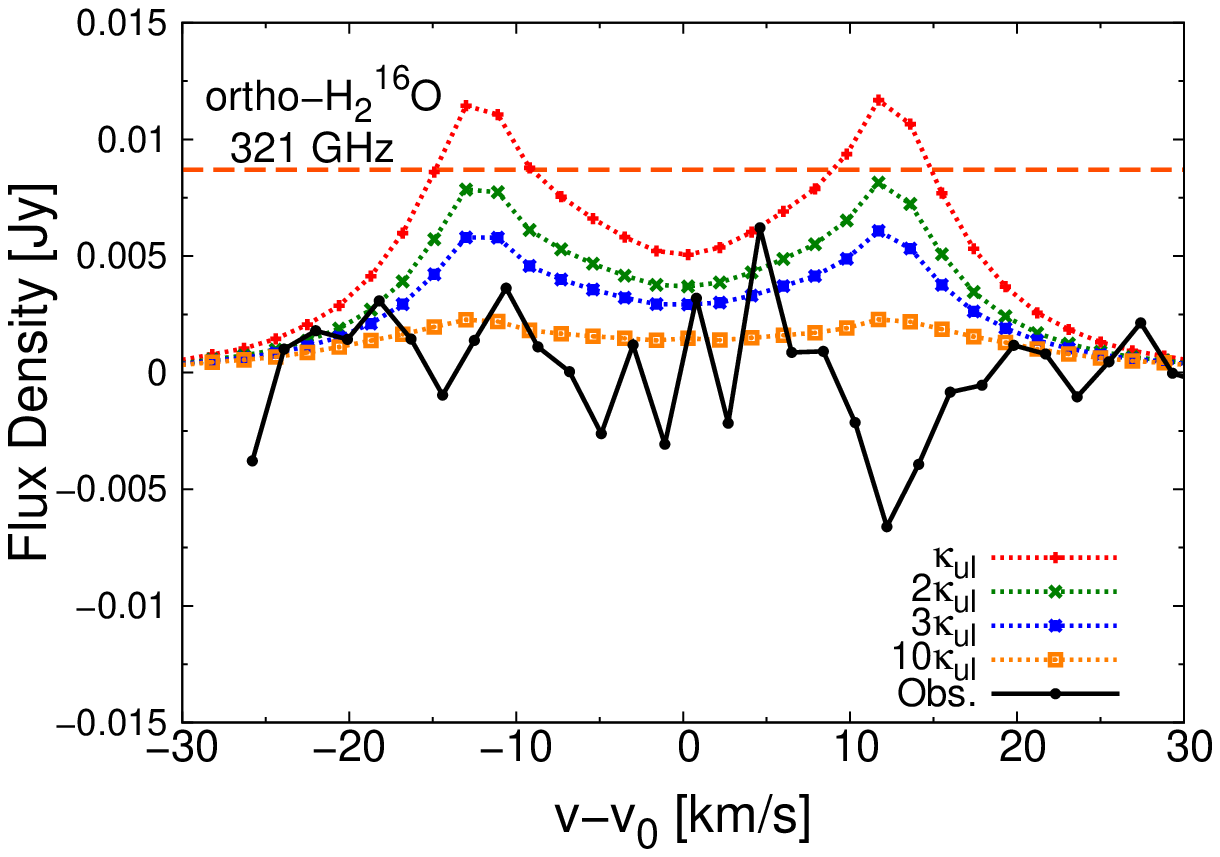}
\includegraphics[scale=0.6]{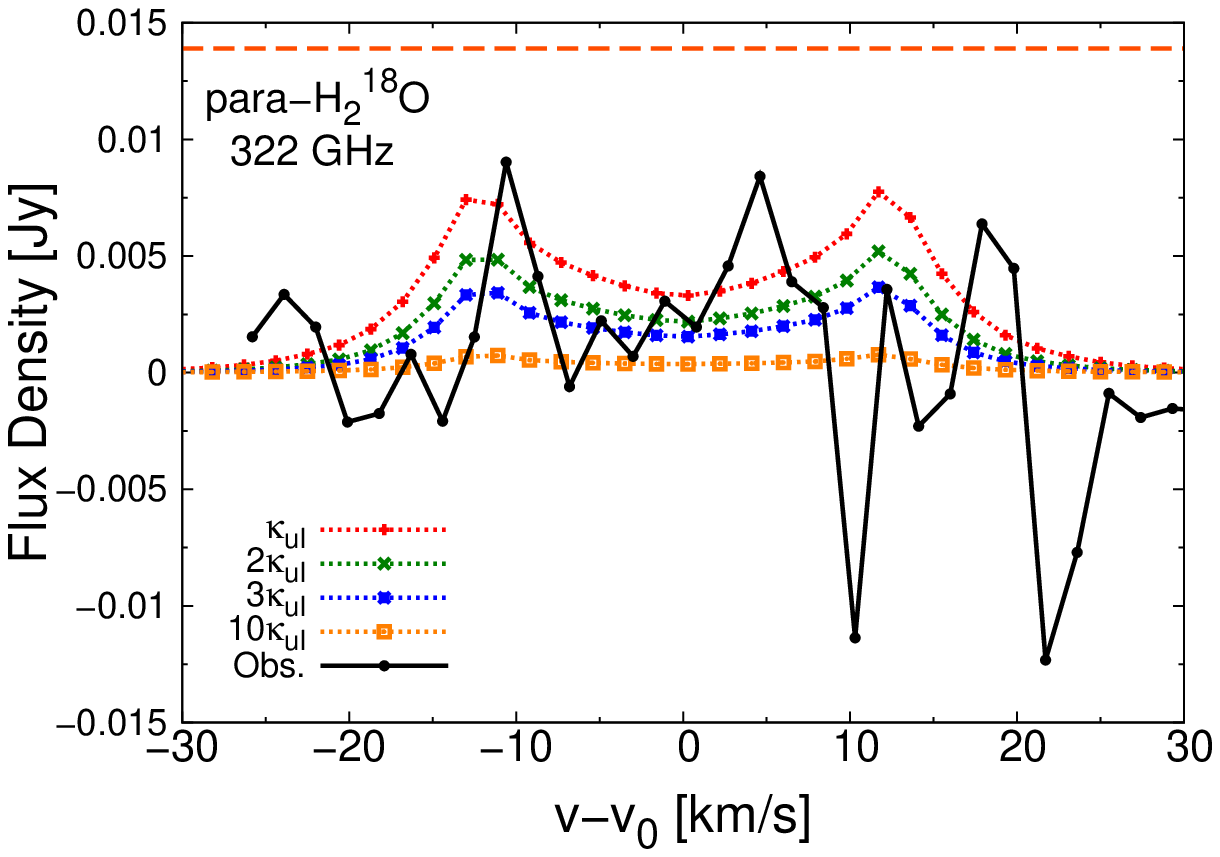}
\end{center}
\vspace{0.55cm}
\caption{
\noindent
The profiles of the ortho-$\mathrm{H_2}$$^{16}\mathrm{O}$ line at 321 GHz (left panel) and the para-$\mathrm{H_2}$$^{18}\mathrm{O}$ line (right panel) at 322 GHz inside 20 au, with a velocity resolution d$v=$1.9 km s$^{-1}$.
The {\it black solid line with circle symbols} is the observed flux density of the disk around HD 163296.
Other lines are the results of our Herbig Ae disk model calculations (see also \citealt{Notsu2017, Notsu2018a}).
The {\it red dotted lines with cross symbols} are the line profiles of our original Herbig Ae disk model.
In the line profiles with {\it green dotted lines with cross symbols}, {\it blue dotted lines with filled square and cross symbols}, and {\it orange dotted lines with square symbols}, we set the values of dust opacity $\kappa_{ul}$ 2, 3, and 10 times larger than our original value.
The horizontal red dashed lines show the values of observed 3$\sigma$ peak flux densities around line centers (see also Table 1).
\vspace{0.5cm}
}\label{Figure2_paperIV}
\end{figure*} 
\begin{figure*}[htbp]
\begin{center}
\includegraphics[scale=0.6]{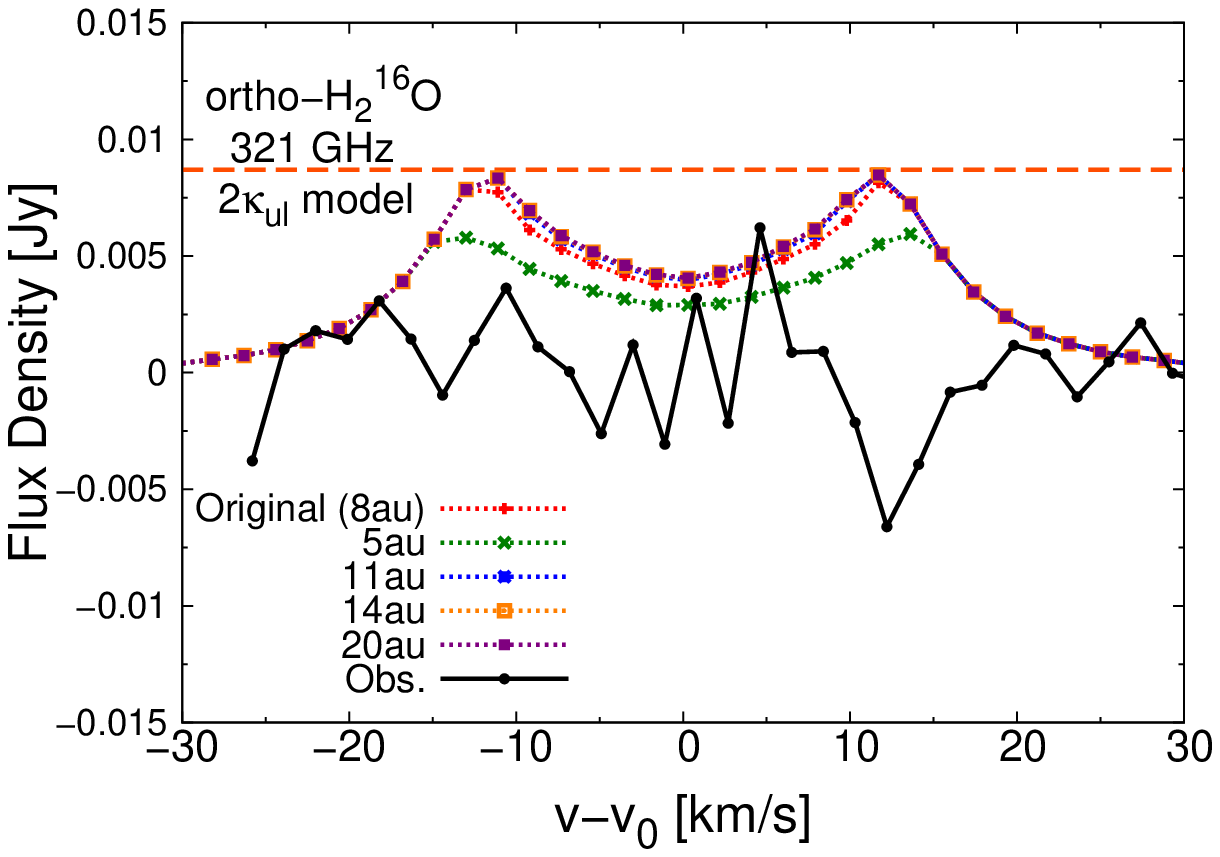}
\includegraphics[scale=0.6]{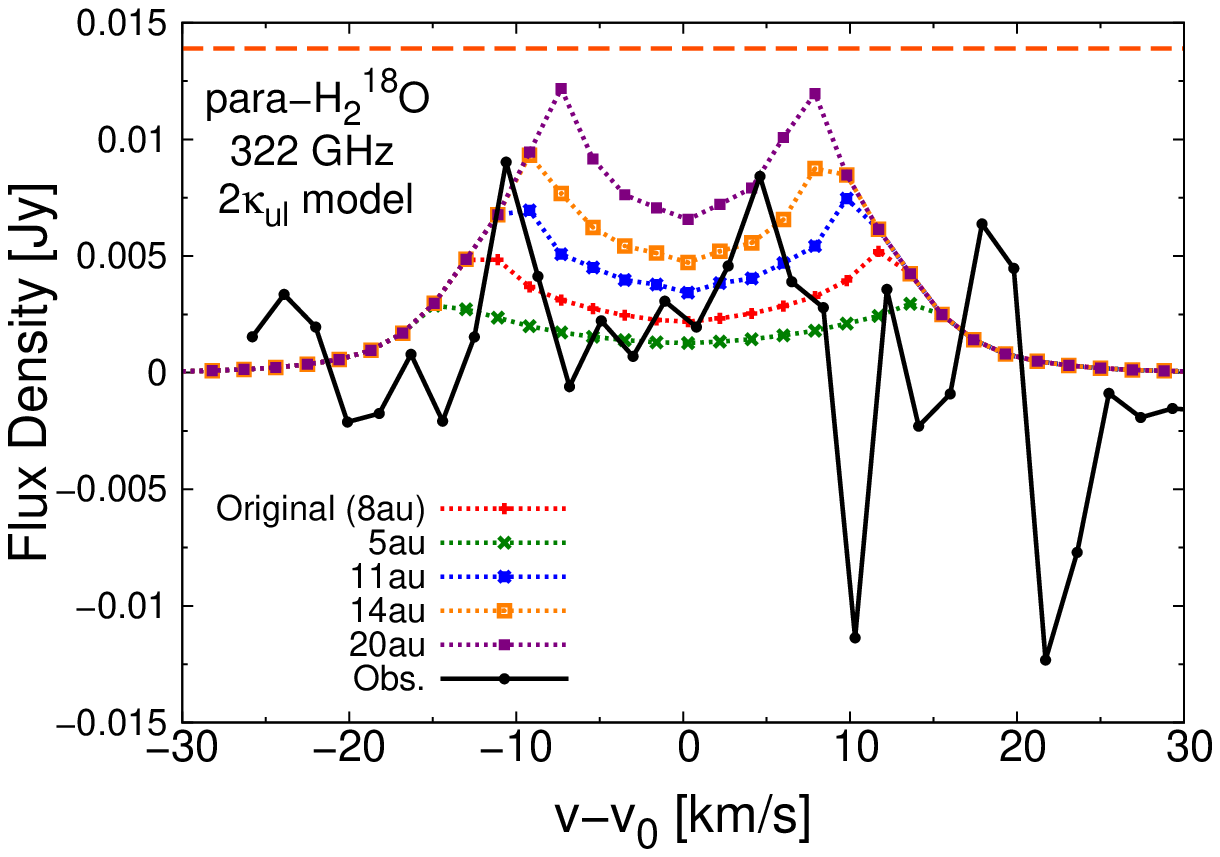}
\end{center}
\vspace{0.55cm}
\caption{
\noindent
The profiles of the ortho-$\mathrm{H_2}$$^{16}\mathrm{O}$ line at 321 GHz (left panel) and the para-$\mathrm{H_2}$$^{18}\mathrm{O}$ line (right panel) at 322 GHz inside 20 au, with a velocity resolution d$v=$1.9 km s$^{-1}$.
The {\it black solid line with circle symbols} is the observed flux density of the disk around HD 163296.
Other lines are the line profiles which are obtained by our Herbig Ae disk model calculations (see also \citealt{Notsu2017, Notsu2018a}).
The {\it red dotted lines with cross symbols} are the line profile with our original water vapor abundance distributions. 
In other line profiles, we artificially change the outer edge of the region with a high $\mathrm{H_2O}$ water vapor abundance ($\sim10^{-5}-10^{-4}$) region to 5 au ({\it green dotted lines with cross symbols}), 11 au ({\it blue dotted lines with filled square and cross symbols}), 14 au ({\it orange dotted lines with perforated square symbols}), and 20 au ({\it purple dotted lines with square symbols}).
We set the values of dust opacity two times larger than our original value (see also Figure \ref{Figure2_paperIV}).
The horizontal red dashed lines show the values of observed 3$\sigma$ peak flux densities around line centers (see also Table 1).
\vspace{0.5cm}
}\label{Figure3_paperIV}
\end{figure*} 
\begin{figure*}[htbp]
\begin{center}
\includegraphics[scale=0.6]{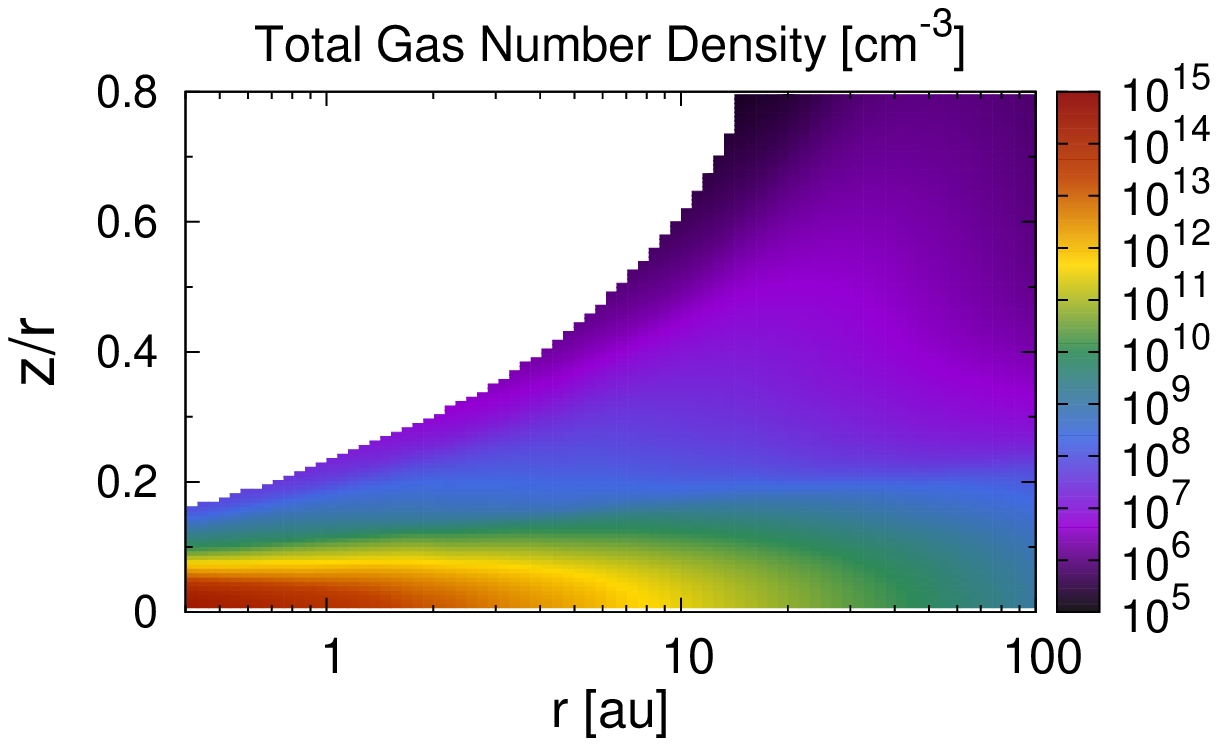}
\includegraphics[scale=0.6]{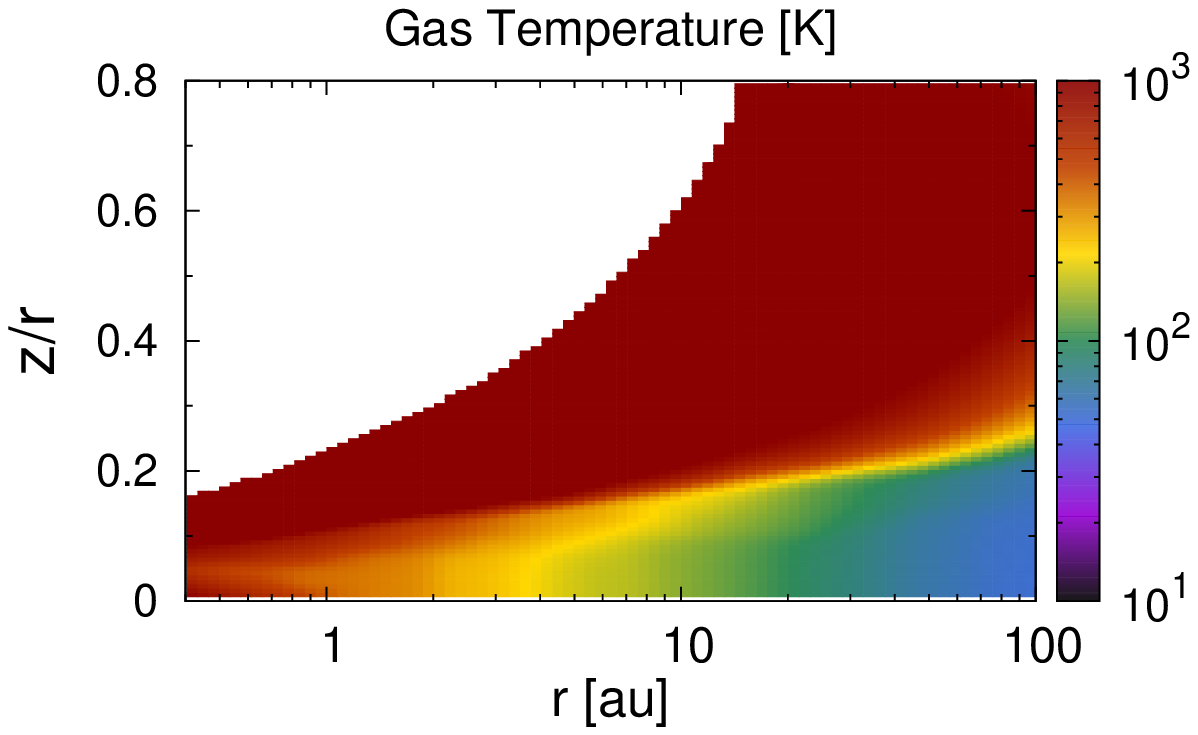}
\includegraphics[scale=0.6]{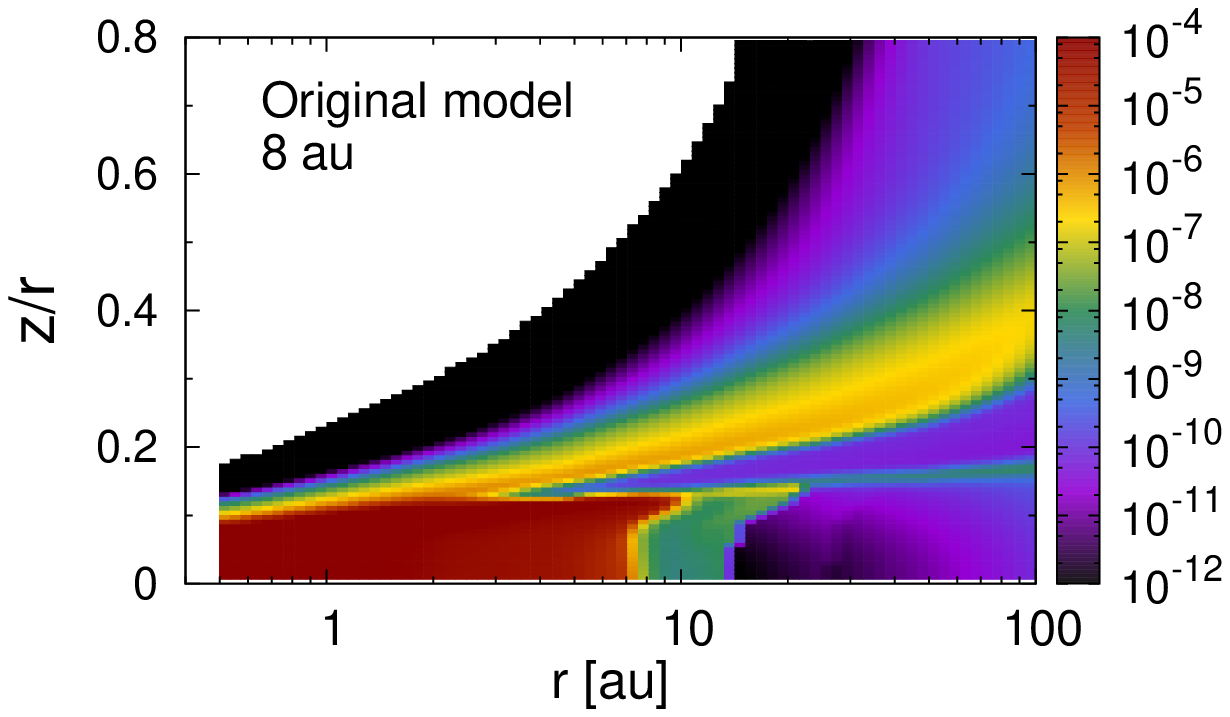}
\end{center}
\vspace{0.4cm}
\caption{\noindent 
The total gas number density in cm$^{-3}$ (top left panel), the gas temperature in K (top right panel), and fractional abundance (relative to total hydrogen nuclei density) distribution of water gas (bottom panel) of a disk around a Herbig Ae star as a function of the disk radius in au and height (scaled by the radius, $z/r$) up to a maximum radius of $r =100$ au.
\vspace{0.5cm}
}\label{Figure4_add_paperIV}
\end{figure*} 
\begin{figure*}[htbp]
\begin{center}
\includegraphics[scale=0.6]{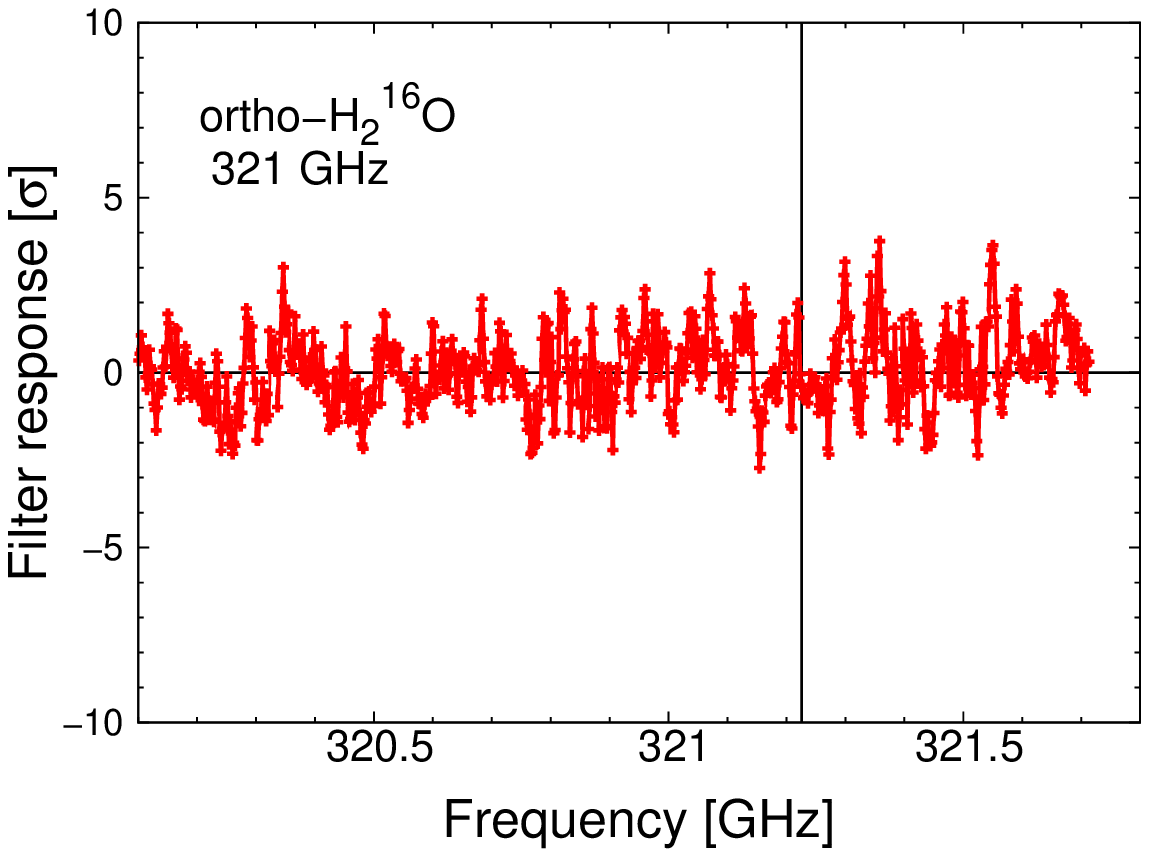}
\includegraphics[scale=0.6]{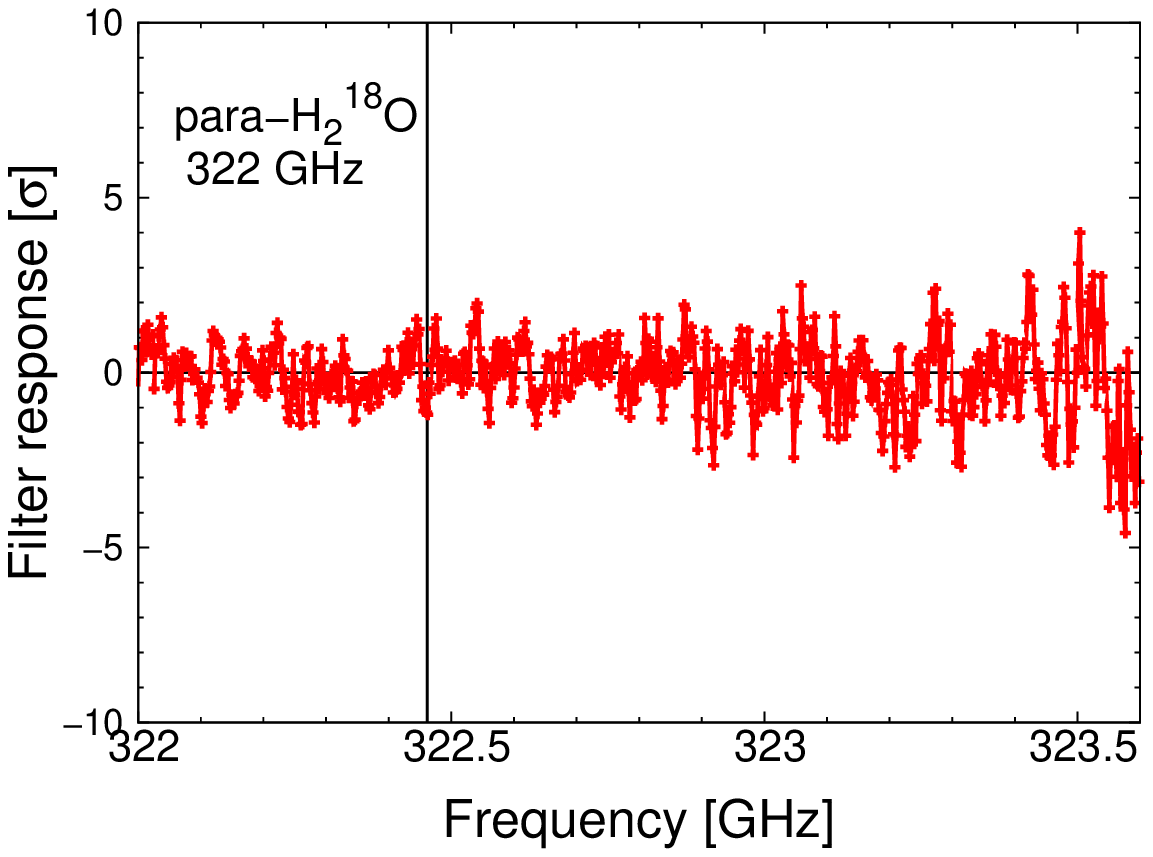}
\includegraphics[scale=0.6]{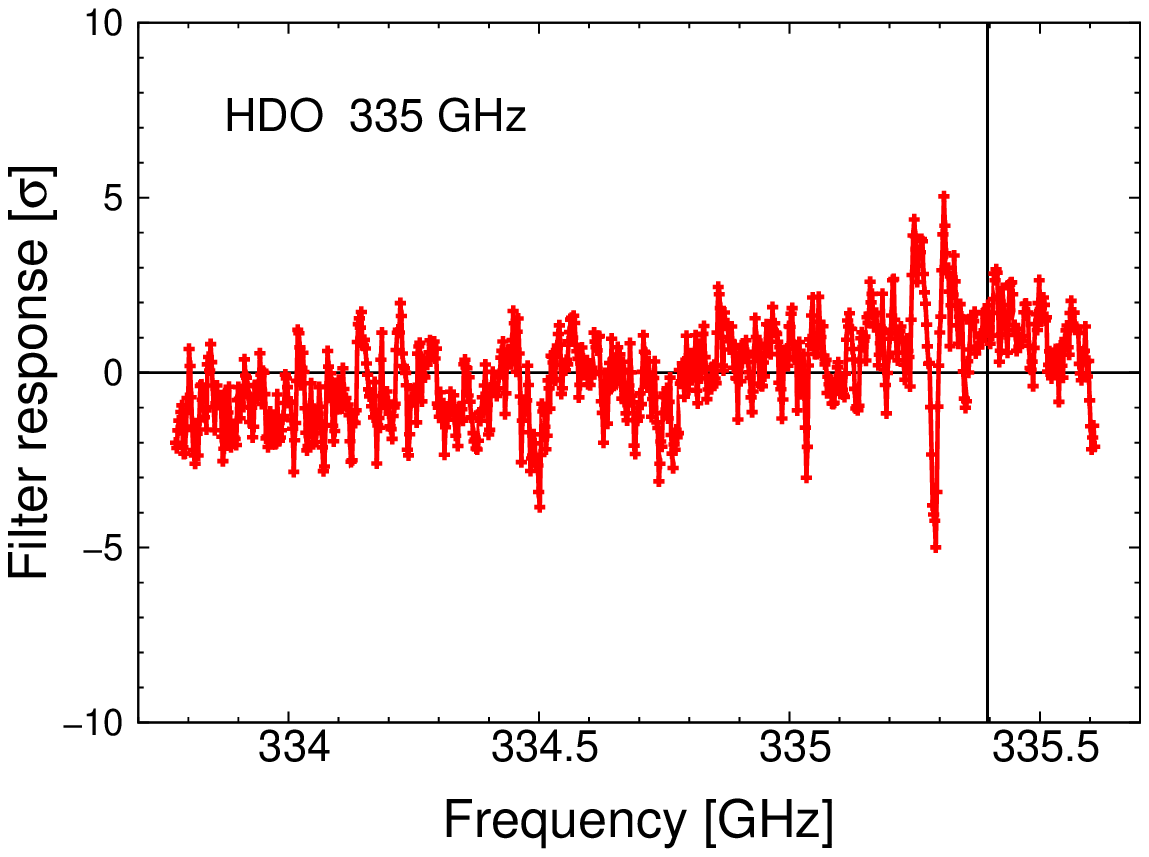}
\end{center}
\vspace{0.6cm}
\caption{
\noindent
The matched filter response functions of the ortho-$\mathrm{H_2}$$^{16}\mathrm{O}$ line at 321 GHz (top left panel), the para-$\mathrm{H_2}$$^{18}\mathrm{O}$ line (top right panel) at 322 GHz, and HDO 335 GHz (bottom panel).
The filter response is normalized by its standard deviation, $\sigma$ (see also \citealt{Loomis2018}).
\vspace{0.5cm}
}\label{Figure4_paperIV}
\end{figure*} 
\noindent
Figure \ref{Figure1_paperIV} shows the observed 
flux densities around the line centers
of the ortho-$\mathrm{H_2}$$^{16}\mathrm{O}$ line at 321 GHz, the para-$\mathrm{H_2}$$^{18}\mathrm{O}$ line at 322 GHz, and the HDO line at 335 GHz of the disk inside 20 au around HD 163296, with a velocity resolution d$v=$1.9 km s$^{-1}$.
The detailed parameters, such as transition quantum numbers ($J_{K_{a}K_{c}}$), wavelength $\lambda$, frequency, $A_{\mathrm{ul}}$, $E_{\mathrm{up}}$, 3$\sigma$\footnote[3]{the σ value is the root-mean-square value of peak flux density} peak flux densities, and 3$\sigma$ total fluxes of lines are listed in Table 1.
These lines have not been detected, 
and we conduct our subsequent analyses using the extracted upper limits for the line fluxes.
In calculating the upper limits, we assume that the velocity width of the double peaked profiles is 30 km s$^{-1}$, on the basis of the velocity width of model calculated line profiles (see Figures \ref{Figure2_paperIV} and \ref{Figure3_paperIV}). 
Depending on the shapes of model line profiles, the actual velocity widths between the line peaks could be smaller than 30 km s$^{-1}$, especially in the cases with large snowline positions. Therefore, in these cases the upper limit values of line fluxes would be over-estimated (see Figures \ref{Figure2_paperIV} and \ref{Figure3_paperIV}, and Table 1).
%
\\ \\
In Figures \ref{Figure2_paperIV} and \ref{Figure3_paperIV}, we compare the upper limit fluxes with the values from our model water line calculations with dust emission \citep{Notsu2015, Notsu2016, Notsu2017, Notsu2018a}, 
to constrain the dust opacity and the line emitting region, respectively, 
and to estimate the necessary observation time for a future clear detection. We calculate the model line profiles with different 
dust opacity values and different outer edges of the region with a high water vapor abundance. 
When we calculate these model line profiles, we include both dust and gas emission components, and we subtract the dust continuum emission component (the values of the calculated fluxes at $v-v_{0}=\pm\infty$) to show the line emission component more clearly.
\\ \\
In our models, we first adopted the physical model of a steady, axisymmetric Keplerian accretion disk with a viscous parameter $\alpha$=$10^{-2}$, a mass accretion rate $\dot{M}$=$10^{-8}M_{\bigodot}$ yr$^{-1}$, and gas-to-dust mass ratio $g/d=100$ surrounding a Herbig Ae star 
with stellar mass $M_{\mathrm{*}}$=2.5$M_{\bigodot}$, stellar radius $R_{\mathrm{*}}$=2.0$R_{\bigodot}$, and effective temperature $T_{\mathrm{*}}$=10,000K \citep{NomuraMillar2005, Nomura2007, Walsh2015}. 
The top panels of Figure \ref{Figure4_add_paperIV} show the total gas number density in cm$^{-3}$ (top left panel) and the gas temperature in K (top right panel) of a disk around a Herbig Ae star (see also Figure 1 of \citealt{Notsu2017}).
Next we calculated the water gas and ice distributions in the disk using chemical kinetics.
The large chemical network \citep{Woodall2007, Walsh2010, Walsh2012, Walsh2015} we use to calculate the disk molecular abundances includes gas-phase reactions and gas-grain interactions (freeze-out, and thermal and nonthermal desorption).
The bottom panel of Figure \ref{Figure4_add_paperIV} shows the fractional abundance (relative to total hydrogen nuclei density) distribution of water gas of the disk (see also Section 3.1 and Figure 2 of \citealt{Notsu2017}).
We found that the water abundance is high (up to $10^{-4}$) in the inner region with higher temperature ($\sim 170$K) within $\sim7-8$ au, and relatively high ($10^{-8}$) between $7-8$ au and 14 au (at the position of the $\mathrm{H_2O}$ snowline, $\sim 120$K) near the equatorial plane. In addition, it is relatively high ($\sim10^{-8}-10^{-7}$) in the hot surface layer and the photodesorbed region of the outer disk, compared to its value ($10^{-12}$) in the regions outside the $\mathrm{H_2O}$ snowline near the equatorial plane. 
Using these data, we calculated the profiles of water emission lines. For more details, see e.g., \citet{Notsu2015, Notsu2016, Notsu2017, Notsu2018a}.
\\ \\
For the calculation of line profiles, we modified the 1D code RATRAN\footnote[4]{\url{http://home.strw.leidenuniv.nl/~michiel/ratran/}} \citep{Hogerheijde2000}.
The data for the line parameters are adopted from the Leiden Atomic and Molecular Database LAMDA\footnote[5]{\url{http://home.strw.leidenuniv.nl/~moldata/}} 
\citep{Schoier2005} for the $\mathrm{H_2}$$^{16}\mathrm{O}$ lines and from the HITRAN Database\footnote[6]{\url{http://www.hitran.org}} (e.g., \citealt{Rothman2013}) for the $\mathrm{H_2}$$^{18}\mathrm{O}$ lines.
Here we note that HDO/H$_{2}$O ratio is considered to be sensitive to the temperature in the disk \citep{vanDishoeck2014}. However, deuterium chemistry is not included in our chemical network, thus we only calculated the $\mathrm{H_2}$$^{16}\mathrm{O}$ and $\mathrm{H_2}$$^{18}\mathrm{O}$ line profiles.
We set the ortho-to-para ratio (OPR) of water to its high-temperature value of 3 throughout the disk \citep{Mumma1987, Hama2013, Hama2016, Hama2018}.
We set the isotope ratio of oxygen $^{16}$O/$^{18}$O to 560 throughout the disk, as \citet{Jorgensen2010} and \citet{Persson2012} adopted.
This $^{16}$O/$^{18}$O value is determined from the observations of the local interstellar medium \citep{Wilson1994}.
We do not include emission from jet components and disk winds in calculating the line profiles. 
\\ \\
First, we fix the location of the abundant water vapor region inside the $\mathrm{H_2O}$ snowline and change the dust opacity.
The original position of the outer edge of the water vapor abundant region is 8 au, and that of the water snowline is 14 au.
In Figure \ref{Figure2_paperIV}, we plot the model profiles of the ortho-$\mathrm{H_2}$$^{16}\mathrm{O}$ 321 GHz line and the para-$\mathrm{H_2}$$^{18}\mathrm{O}$ 322 GHz line with four dust opacity values.
The {\it red dotted lines with cross symbols} are the line profile of our original Herbig Ae model \citep{Notsu2017, Notsu2018a}. In this model, the value of mm dust opacity
$\kappa_{\mathrm{mm}}$ is 1.0 cm$^{2}$ g$^{-1}$.
In the line profiles with {\it green dotted lines with cross symbols}, {\it blue dotted lines with filled square and cross symbols}, and {\it orange dotted lines with perforated square symbols}, we set the values of dust opacity 2, 3, and 10 times larger, respectively, than our original value in order to investigate the influence of dust opacity on line properties.
The calculated line fluxes are listed in Table 2.
We note that the dust opacity at sub-millimeter wavelengths can vary by a factor of around 10, depending on the properties of the dust grains (e.g., \citealt{Miyake1993, Draine2006}).
In our fiducial disk model, dust opacities appropriate for the dark cloud model are used and are relatively small at sub-millimeter wavelengths, compared to models with grain growth (see e.g., \citealt{NomuraMillar2005, Aikawa2006}, and paper I).
In our model calculations, the emission of our observed Band 7 water lines is optically thick in the disk midplane within the $\mathrm{H_2O}$ snowline. In contrast, the sub-millimeter dust emission is marginally optically thick. The dust emission becomes stronger as the value of the dust opacity becomes larger, and the line emission becomes obscured by the dust emission.
As a result, the apparent line intensities obtained by subtracting dust continuum emission become smaller (for more details, see \citealt{Notsu2017, Notsu2018a}).
 \\ \\
Here we note that the dust properties are important because they affect the physical and chemical structure of protoplanetary disks (for details, see, e.g., \citealt{Nomura2007}).
The total surface area of dust grains has an influence on the abundances of molecules through determining the gas and ice balance.
In addition, since dust grains are the dominant opacity source in the disks, they determine the temperature profiles and the UV radiation field throughout the disk.
As the location of the $\mathrm{H_2O}$ snowline is sensitive to the temperature, it strongly depends on the dust opacity especially at mid-infrared 
wavelength because the peak wavelength of the blackbody radiation
from the disk midplane around the $\mathrm{H_2O}$ snowline ($\sim 100-200$K) is mainly around $10-30 \mu$m. 
The sub-millimeter dust opacity, on the other hand, is not a direct indicator of the dust properties which affect the location of the $\mathrm{H_2O}$ snowline.
\\ \\
According to Figure \ref{Figure2_paperIV} and Tables 1 and 2, the observed 3$\sigma$ upper limit peak flux density of the para-$\mathrm{H_2}$$^{18}\mathrm{O}$ 322 GHz line is larger than model calculated value with original dust opacity, and that of the ortho-$\mathrm{H_2}$$^{16}\mathrm{O}$ 321 GHz line is close to the calculated value for the model with two times larger dust opacity.
Here we note that in the cases of our target water lines, most of the line emission comes from the region with a high water gas abundance ($\gtrsim10^{-5}$) in the Herbig Ae disk \citep{Notsu2017, Notsu2018a}.
Therefore, we consider that the mm dust opacity $\kappa_{\mathrm{mm}}$ is larger than 2.0 cm$^{2}$ g$^{-1}$ to explain the water line upper limit, if the outer edge of the water vapor abundant region and also the position of the water snowline is located at or beyond 8 au (see also Figure \ref{Figure3_paperIV}).
Previous dust continuum observations of this object (e.g., \citealt{Boneberg2016}) show that $\kappa_{\mathrm{mm}}$$\sim$3.0 cm$^{2}$ g$^{-1}$.
In the cases with three times larger dust opacity, the model peak flux densities correspond to around 2$\sigma$ for ortho-$\mathrm{H_2}$$^{16}\mathrm{O}$ 321 GHz line and around 1$\sigma$ for para-$\mathrm{H_2}$$^{18}\mathrm{O}$ 322 GHz line. 
Therefore, the observation time executed (20\% of our proposed time in our Cycle 3 proposal) was not enough to test our model.
Here we note that the dust optical depth $\tau_{d}$ is $\sim$0.2 at r$\sim$5 au for the model with the original dust opacity ($\kappa_{\mathrm{mm}}=1.0$ cm$^{2}$ g$^{-1}$). 
Meanwhile, previous ALMA and VLA observations with lower
spatial resolution suggests $\tau_{d}$$\sim$0.55 at r$<$50 au \citep{Guidi2016}, which is a few times larger than our original value at r$\sim$5 au.
\\ \\
In Figure \ref{Figure3_paperIV}, we fix the dust opacity and artificially change the outer edge of the region with a high $\mathrm{H_2}\mathrm{O}$ vapor abundance ($10^{-5}$) from 8 au (\citealt{Notsu2017, Notsu2018a}, see also Figure \ref{Figure4_add_paperIV} of this paper), to 5 au ($T_{g}\sim$180K), 11 au ($T_{g}\sim$135K), 14 au ($T_{g}\sim$120K, the exact water snowline position), and 20 au ($T_{g}\sim$100K).
Figure \ref{Figure8_add_paperIV} in the Appendix of this paper shows the artificially changed fractional abundance distributions of water vapor.
We set the values of dust opacity two times larger than that of our original Herbig Ae model ($\kappa_{\mathrm{mm}}=2.0$ cm$^{2}$ g$^{-1}$, see also Table 2 and Figure \ref{Figure2_paperIV}).
The calculated line fluxes are listed in Table 3.
As the region with a high water gas abundance becomes larger, the flux density of the line peaks becomes larger, and the width between the two line peaks becomes narrower. It is because the Keplerian velocity in the outer disk is smaller than that in the inner disk.
Since the $E_{\mathrm{up}}$ of para-$\mathrm{H_2}$$^{18}\mathrm{O}$ 322 GHz line (467.9K) is smaller than that of the ortho-$\mathrm{H_2}$$^{16}\mathrm{O}$ 321 GHz line (1861.2K), the former line profile is expected to be more sensitive to the change of position of the $\mathrm{H_2O}$ snowline. It is because the temperature around the $\mathrm{H_2O}$ snowline is around $100-200$K, and thus the line emitting region of the former line extends around the $\mathrm{H_2O}$ snowline, in contrast that of the latter line is localized more inward (for more details, see \citealt{Notsu2017, Notsu2018a}).
According to the right-hand panel of Figure \ref{Figure3_paperIV}, the position of the water snowline will be inside 20 au, if the mm dust opacity $\kappa_{\mathrm{mm}}$ is 2.0 cm$^{2}$ g$^{-1}$ (see also Figure \ref{Figure2_paperIV}). 
If $\kappa_{\mathrm{mm}}$ is larger than 2.0 cm$^{2}$ g$^{-1}$, the outer edge of the region with a high $\mathrm{H_2}\mathrm{O}$ vapor abundance the position of $\mathrm{H_2O}$ snowline can be larger than 20 au.
\\ \\
As we explained in Section 2.1, the integration time on source is 0.723 hours, which is around 20\% of the requested time in our Cycle 3 proposal.
Future observations of sub-millimeter water lines with longer observation time are required to confine the water line fluxes and the position of the $\mathrm{H_2O}$ snowline in the disk midplane in detail for the disk around HD 163296.
Following the approach undertaken by \citet{Carney2019} in their discussion of CH$_{3}$OH, we estimate
the additional observation times required for the detections of water lines in the disk around HD 163296 with ALMA.
The values of peak flux densities from our model calculations suggest that to obtain significant detection (5$\sigma$) of ortho-$\mathrm{H_2}$$^{16}\mathrm{O}$ 321 GHz line, around 3 times longer integration time would be needed if $\kappa_{\mathrm{mm}}$ is 2.0 cm$^{2}$ g$^{-1}$
and around 5 times longer integration time (similar to the requested time in our Cycle 3 proposal) would be needed if $\kappa_{\mathrm{mm}}$ is 3.0 cm$^{2}$ g$^{-1}$ where we assume the outer edge of the region with a high $\mathrm{H_2}\mathrm{O}$ vapor abundance ($10^{-5}$) is larger than 8 au.
Moreover, to obtain 3$\sigma$ and 5$\sigma$ detections of para-$\mathrm{H_2}$$^{18}\mathrm{O}$ 322 GHz line, around 6 and 16 times longer integration time would be needed, respectively, if $\kappa_{\mathrm{mm}}$ is 2.0 cm$^{2}$ g$^{-1}$ and the outer edge of the region with a high $\mathrm{H_2}\mathrm{O}$ vapor abundance ($10^{-5}$) is at 8 au. 
In these time estimations, we assume the similar observational conditions, such as weather and numbers of antenna, to our previous observations.
\\ \\
Here we note that \citet{Carr2018} reported tentative detection (2-3$\sigma$) of these two water lines with ALMA
toward the disk around a T Tauri star, AS 205N,
which has high mass accretion rate ($\dot{M}$=$3\times10^{-7}M_{\bigodot}$ yr$^{-1}$). The 3$\sigma$ flux density at 321 GHz of their observation is around 3 mJy.
\subsection{Matched filtering analysis}
\noindent
\citet{Loomis2018} recently proposed a new method to detect the weak line emissions from Keplerian rotating disks using observed visibility data (matched filter analysis).
In this method, first we generated a Keplerian filter in the image plane with the same position and inclination angles of the source disk. 
The matched filter tool VISIBLE \footnote[7]{freely available at \url{https://github.com/AstroChem/VISIBLE}} then cross-correlates the transformation of this filter in the $uv$ plane with the visibility data points from our observation  (see also \citealt{Carney2017, Carney2018, Carney2019, Booth2018a}).
\\ \\
Figure \ref{Figure4_paperIV} shows the matched filter response functions of the ortho-$\mathrm{H_2}$$^{16}\mathrm{O}$ 321 GHz line (top left panel), the para-$\mathrm{H_2}$$^{18}\mathrm{O}$ 322 GHz line (top right panel), and HDO 335 GHz line (bottom panel). 
The filter response is normalized by its standard deviation, $\sigma$ (see also \citealt{Loomis2018}).
We confirm the non-detection of all three lines as also found in the image-plane analyses.
\section{Dust continuum image and radial profiles}
\begin{figure*}[htbp]
\begin{center}
\includegraphics[scale=0.4]{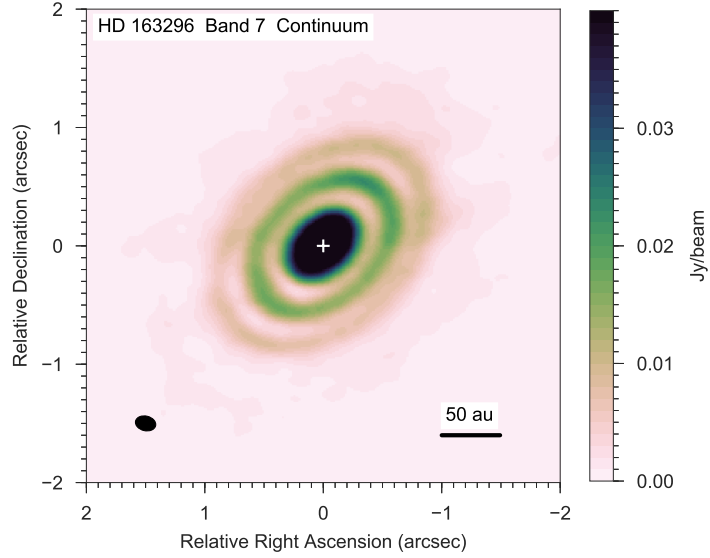}
\end{center}
\caption{
\noindent
ALMA continuum image of the disk around HD 163296 at 0.9 mm (Band 7).
The black ellipse at the bottom left corner shows the synthesized beam ($0.174\times0.124$ arcsec).
The black solid line at the bottom right corner shows the linear scale of 50 au in this disk.
}\label{Figure5_paperIV}
\end{figure*} 
\begin{figure*}[htbp]
\begin{center}
\includegraphics[scale=0.65]{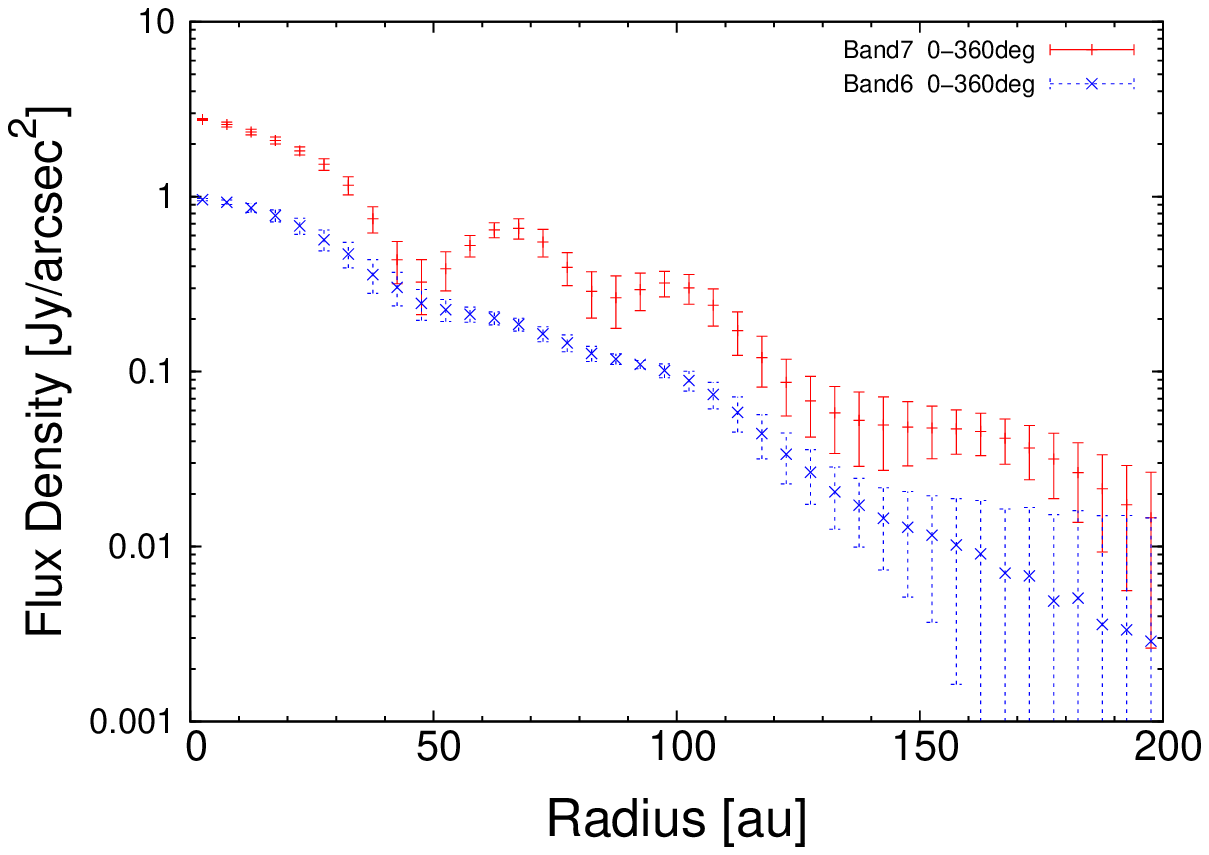}
\includegraphics[scale=0.65]{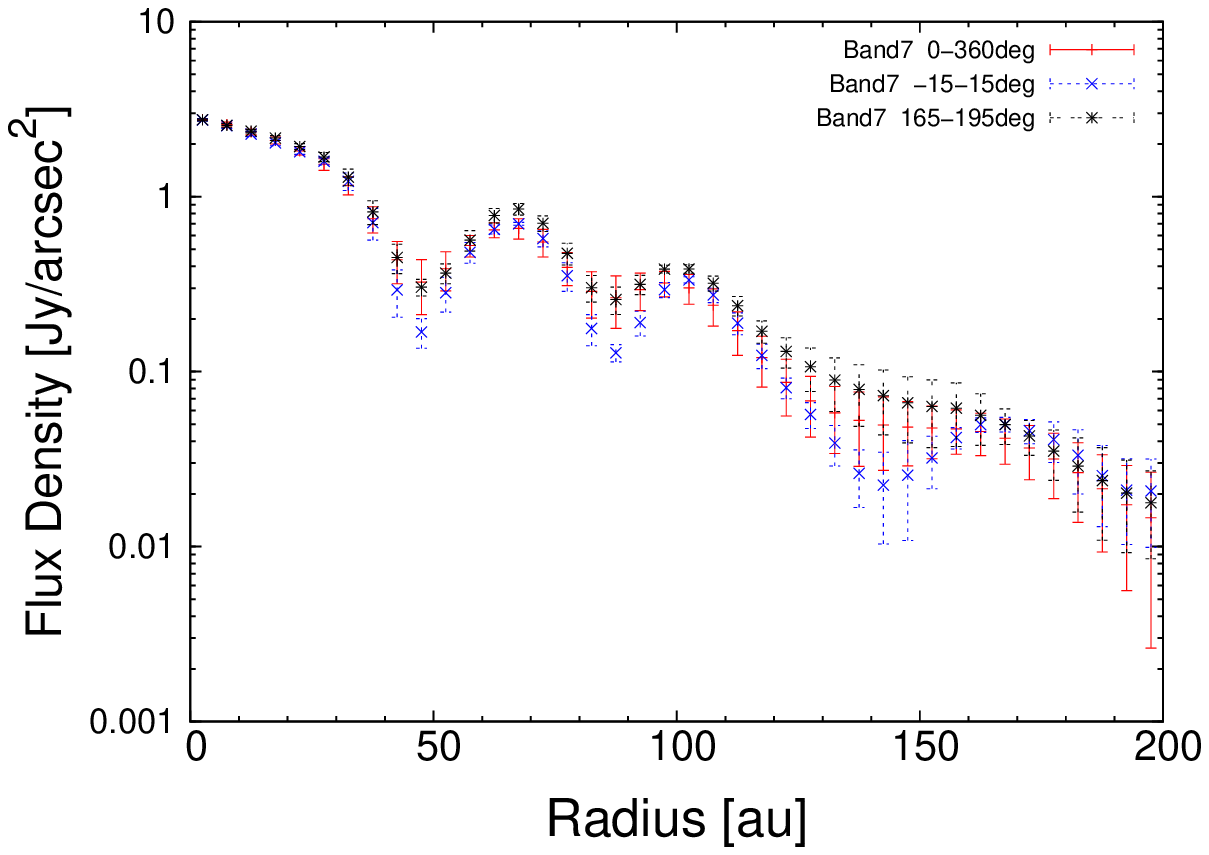}
\end{center}
\vspace{0.6cm}
\caption{
\noindent (Left panel): Radial profiles of the flux densities in Jy/$\mathrm{arcsec}^{2}$ averaged over full azimuthal angle (0-360 $\deg$) for Band 7 ({\it red cross points}) and Band 6 ({\it blue cross points}).
(Right panel): Band 7 radial profiles of the flux densities in Jy/$\mathrm{arcsec}^{2}$ averaged over 0-360 $\deg$ ({\it red cross points}), $-15\sim15 \deg$ ({\it blue cross points}), and $-165\sim195 \deg$ ({\it black asterisk points}).
The error bar is determined from the standard deviation through the azimuthal averaging.
We determine the 0 $\deg$ as the south-east direction of the major axis.}
\label{Figure6_paperIV}
\end{figure*} 
\begin{deluxetable*}{rrrrrrrrr}
\tablewidth{0pt}
\tablecaption{Parameters and observed upper limits of our target water lines in ALMA Band 7 \ \label{tab:5-T1}}
\tablehead{
\colhead{isotope}&
\colhead{$J_{K_{a}K_{c}}$}&
\colhead{\ $\lambda$}&
\colhead{\ Frequency}& \colhead{\ $A_{\mathrm{ul}}$}&
\colhead{$E_{\mathrm{up}}$} & \colhead{3$\sigma$ peak flux density\tablenotemark{a}} & \colhead{3$\sigma$ total flux\tablenotemark{a,b}} \\
\colhead{}&\colhead{}&\colhead{[$\mu$m]}&
\colhead{[GHz]}& \colhead{[s$^{-1}$]}&
\colhead{[K]}& \colhead{[mJy]}& \colhead{[W $\mathrm{m}^{-2}$]}}
\startdata     
 ortho-$\mathrm{H_2}$$^{16}\mathrm{O}$ & 10$_{29}$-9$_{36}$ & 933.2767 & 321.22568 & 6.17$\times10^{-6}$ & 1861.2 &$<$ 8.7 \ \ \ \ \ \ \ \ \ \ \ \ & $<$ 5.3$\times$$10^{-21}$ \\  
 para-$\mathrm{H_2}$$^{18}\mathrm{O}$ & 5$_{15}$-4$_{22}$ & 929.6894 & 322.46517 & 1.06$\times10^{-5}$ & 467.9 &$<$ 13.9 \ \ \ \ \ \ \ \ \ \ \ \ & $<$ 8.5$\times$$10^{-21}$\\
      HDO \ \ \ \ \ &3$_{31}$-4$_{22}$ & 893.8476 & 335.39550 & 2.61$\times$$10^{-5}$ & 335.3 &$<$ 7.3 \ \ \ \ \ \ \ \ \ \ \ \  & $<$ 6.3$\times$$10^{-21}$\\
\enddata
\tablenotetext{a}{In calculating the upper limit values of peak flux densities and total fluxes, we integrate the line flux components within 20 au (circular aperture) from the central star. In addition, the $\sigma$ value is the root-mean-square value of peak flux density.}
\tablenotetext{b}{In calculating the upper limit values of total fluxes, we set the velocity width of the double peaked profiles as 30 km s$^{-1}$, according to the velocity width of model calculated line profiles (see Figures \ref{Figure2_paperIV} and \ref{Figure3_paperIV}).}
\end{deluxetable*}
\begin{deluxetable*}{rrrrrrrrrr}
\tablewidth{0pt}
\tablecaption{Calculated peak flux densities and total fluxes of our target water lines in ALMA Band 7 with different values of dust opacity \ \label{tab:5-T2}}
\tablehead{
\colhead{isotope}&
\colhead{\ Frequency}&
\colhead{}&
\colhead{Peak flux}&
\colhead{density\tablenotemark{a,b}} &
\colhead{[mJy]} &
\colhead{}&
\colhead{Total flux\tablenotemark{a,b}}&
\colhead{[W $\mathrm{m}^{-2}$]} &
\colhead{} \\
\colhead{}&
\colhead{[GHz]}&
\colhead{$\kappa_{\mathrm{ori}}$}&
\colhead{2$\kappa_{\mathrm{ori}}$} &
\colhead{\ \ \ \ 3$\kappa_{\mathrm{ori}}$} &
\colhead{10$\kappa_{\mathrm{ori}}$} &
\colhead{$\kappa_{\mathrm{ori}}$}&
\colhead{2$\kappa_{\mathrm{ori}}$}&
\colhead{3$\kappa_{\mathrm{ori}}$} &
\colhead{10$\kappa_{\mathrm{ori}}$} 
}
\startdata      
ortho-$\mathrm{H_2}$$^{16}\mathrm{O}$ & 321.22568 &12.6&8.9 \ \ \ \ \ &6.6 \ \ \ &2.5 \ \ \ & 2.4$\times$$10^{-21}$ & 2.4$\times$$10^{-21}$ & 1.3$\times$$10^{-21}$ & 6.4$\times$$10^{-22}$\\  
para-$\mathrm{H_2}$$^{18}\mathrm{O}$ & 322.46517 &8.5&5.7 \ \ \ \ \ &4.1 \ \ \ &0.87 \ \ \ &1.9$\times$$10^{-21}$ & 1.8$\times$$10^{-21}$& 8.4$\times$$10^{-22}$ & 1.9$\times$$10^{-22}$\\
\enddata
\tablenotetext{a}{When we calculate these model line fluxes, we include both dust and gas emission components, and we subtract dust emission components after calculations. We set the four values of dust opacity $\kappa$ (original value, 2, 3, and 10 times larger values). in order to investigate the influence of dust opacity on line properties.}
\tablenotetext{b}{In calculating these model line fluxes, we integrate the flux components within 20 au (circular aperture) from the central star.}
\end{deluxetable*}
\begin{deluxetable*}{rrrrrrrrrrrr}
\tablewidth{0pt}
\tablecaption{Calculated peak flux densities and total fluxes of our target water lines in ALMA Band 7 with different outer edge of high water vapor abundance region \ \label{tab:5-T3}}
\tablehead{
\colhead{isotope}&
\colhead{\ Frequency}&
\colhead{}&
\colhead{Peak flux} &
\colhead{density\tablenotemark{a,b}} &
\colhead{[mJy]} &
\colhead{}&
\colhead{}&
\colhead{Total Flux\tablenotemark{a,b}}&
\colhead{[W $\mathrm{m}^{-2}$]} &
\colhead{}&
\colhead{} \\
\colhead{}&
\colhead{[GHz]}&
\colhead{5 au}&
\colhead{8 au} &
\colhead{\ \ \ \ 11 au} &
\colhead{14 au} &
\colhead{20 au} &
\colhead{5 au}&
\colhead{8 au} &
\colhead{11 au} &
\colhead{14 au} &
\colhead{20 au}
}
\startdata      
ortho-$\mathrm{H_2}$$^{16}\mathrm{O}$ & 321.22568 &5.9&8.9 \ \ \ \ \ &9.1 \ \ &9.2 \ \ &9.2 \ \ &2.0$\times$$10^{-21}$&2.4$\times$$10^{-21}$&2.6$\times$$10^{-21}$&2.6$\times$$10^{-21}$&2.6$\times$$10^{-21}$\\  
para-$\mathrm{H_2}$$^{18}\mathrm{O}$ & 322.46517 &3.0&5.7 \ \ \ \ \ &8.7 \ \ &10.9 \ \ &14.4 \ \ &1.1$\times$$10^{-21}$&1.8$\times$$10^{-21}$&2.5$\times$$10^{-21}$&3.0$\times$$10^{-21}$&3.7$\times$$10^{-21}$\\
\enddata
\tablenotetext{a}{When we calculate these model line fluxes, we include both dust and gas emission components, and we subtract dust emission components after calculations. 
We set the values of dust opacity $\kappa$ two times larger than that of our original Herbig Ae model (see also Figure \ref{Figure2_paperIV} and Table 2).
We set the five values of outer edge of high water vapor abundance region in the inner disk (5 au, original value 8 au, 11 au, 14 au, and 20 au). 
}
\tablenotetext{b}{In calculating these model line fluxes, we integrate the flux components within 20 au (circular aperture) from the central star.}
\end{deluxetable*}
%
%
\noindent Figure \ref{Figure5_paperIV} shows the dust continuum emission map of the disk around HD 163296 at ALMA Band 7.
The positions of bright rings and dark gaps in our observed Band 7 data are consistent with those indicated by the previous Cycle 2 observation of the Band 6 dust continuum emission \citep{Isella2016}. The resolution of the Band 7 image is $\sim$1.4 times smaller than that of Band 6 image reported in \citet{Isella2016}, and $\sim$2.4 times smaller than that of our Band 6 image. In addition, \citet{Dent2019} recently reported the Band 7 dust continuum image of this object, and the positions of rings and gaps in dust continuum emissions are consistent in both data.
\\ \\
To confirm the multiple ring and gap structures in detail, we plot the azimuthally-averaged radial profiles of the dust continuum emission in Figure \ref{Figure6_paperIV}.
There are three prominent gaps at around 50, 83 and 137 au, as previously reported in recent observations \citep{Isella2016, Isella2018, Andrews2018, Dent2019}.
The gap depths in Band 7 data appear deeper than those in our Band 6 data because of the difference in spatial resolutions.
\\ \\
These multiple gaps and rings have been also found for several protoplanetary disks, such as HL Tau (e.g., \citealt{ALMA2015, Akiyama2016, Carrasco2016, Pinte2016}) and TW Hya (e.g., \citealt{Akiyama2015, Rapson2015, Andrews2016, Nomura2016, Tsukagoshi2016, vanBoekel2017, Huang2018a}).
Observations from the Disk Substructures at High Angular Resolution Project (DSHARP) recently published show that the continuum substructures are ubiquitous in disks.
The most common substructures are narrow emission rings and depleted gaps, although large-scale spiral patterns and small arc-shaped azimuthal asymmetries are also present in some cases (e.g., \citealt{Andrews2018, Huang2018b, Huang2018c, Isella2018}).
\\ \\
The origins of multiple gap and ring patterns are still debated.
Several theoretical studies proposed that the planet-disk interaction causes material clearance around the orbits of the planets (e.g., \citealt{Kanagawa2015a, Kanagawa2015b, Kanagawa2016, Kanagawa2018, Jin2016, Dong2018}). There are another scenarios to explain these patterns without planets, such as efficient particle growth and fragmentation of dust grains around snowlines (e.g., \citealt{Ros2013, Banzatti2015, Zhang2015, Okuzumi2016, Pinilla2017}), particle trapping at the pressure bump structures in the disk surface density close to the outer edge of the dead-zone (e.g., \citealt{Flock2015, Pinilla2016, Ruge2016}), zonal flows from magneto-rotational instability (e.g., \citealt{Bethune2016}), secular gravitational instability (e.g., \citealt{Takahashi2014, Takahashi2016, Tominaga2017}), or baroclinic instability triggered by dust settling \citep{LorenAguilar2015}.
\\ \\
Current and future theoretical studies and observations of the dust continuum emission and gas emission (e.g., CO lines) at higher angular resolution with longer observation time are required to clarify the origins of these structures (e.g., \citealt{vanderMarel2018}). Here we note that recently \citet{vanderMarel2019} suggested that such gap radii generally do not correspond to the orbital radii of snowlines of common molecules, such as CO, CO$_{2}$, CH$_{4}$, N$_{2}$, and NH$_{3}$, and the planet scenario has possibility to explain the gaps, especially if the disk viscosity is low and the gaps can be explained by Neptune-mass planets.
\section{Conclusions}
\noindent
In this paper, we used ALMA to obtain upper limit fluxes of sub-millimeter ortho-H$_{2}$$^{16}$O 321 GHz, para-H$_{2}$$^{18}$O 322 GHz, and HDO 335 GHz lines from 
the protoplanetary disk around the Herbig Ae star, HD 163296.
The targeted lines are considered to be the prime candidate water lines at sub-millimeter wavelengths to locate the position of the $\mathrm{H_2O}$ snowline.
These lines have not been detected, and we obtained the upper limit values of peak flux densities and total fluxes.
We compared the upper limit fluxes with the values calculated by our model water line calculations with dust emission \citep{Notsu2015, Notsu2016, Notsu2017, Notsu2018a}.
We constrained the line emitting region and the dust opacity from the observations.
We find that the mm dust opacity $\kappa_{\mathrm{mm}}$ is larger than 2.0 cm$^{2}$ g$^{-1}$ to explain the water line properties, 
if the outer edge of the water vapor abundant region and also the position of the water snowline is beyond 8 au.
In addition, the position of the water snowline will be inside 20 au, if the mm dust opacity $\kappa_{\mathrm{mm}}$ is 2.0 cm$^{2}$ g$^{-1}$. 
We also report multiple ring and gap patterns in 0.9 mm (Band 7) dust continuum emission with 15 au resolution. The positions of bright rings and dust depleted dark gaps are consistent with those indicated by the previous observations \citep{Isella2016, Dent2019}.
Future observations of the dust continuum emission at higher angular resolution and sub-millimeter water lines with longer observation time are required to clarify the detailed structures and the position of the $\mathrm{H_2O}$ snowline in the disk midplane.\\ \\
\acknowledgments
\noindent We are grateful to Professor Inga Kamp and Dr. Satoshi Okuzumi for their useful comments.
We thank the referee for many important suggestions and comments.
This paper makes use of the following ALMA data: ADS/JAO.ALMA \#2015.1.01259.S and ADS/JAO.ALMA \#2013.1.00601.S.
ALMA is a partnership of European Southern Observatory (ESO) (representing its member states), National Science Foundation (USA), and
National Institutes of Natural Sciences (Japan), together with National Research Council (Canada), National Science Council and Academia Sinica Institute of
Astronomy and Astrophysics (Taiwan), and Korea Astronomy and Space Science Institute (Korea), in cooperation with the Republic of Chile.
The Joint ALMA Observatory is operated by ESO, Associated Universities, Inc/National Radio Astronomy Observatory (NRAO), and National Astronomical Observatory of Japan.
Our numerical studies were carried out on SR16000 at Yukawa Institute for Theoretical Physics (YITP) and computer systems at Kwasan and Hida Observatory (KIPS) in 
Kyoto University, and PC cluster at Center for Computational Astrophysics, National Astronomical Observatory of Japan.
ALMA Data analysis was carried out on the Multi-wavelength Data Analysis System operated by the Astronomy Data Center (ADC), National Astronomical Observatory of Japan.
This work is supported by JSPS (Japan Society for the Promotion of Science) Grants-in-Aid for Scientific Research (KAKENHI) (Grant Number; 25108004, 25108005, 25400229, 15H03646, 15K17750, 17K05399),
by Grants-in-Aid for JSPS fellows (Grant Number; 16J06887), and by the Astrobiology Center Program of National Institutes of Natural Sciences (NINS) (Grant Number; AB281013).
CW acknowledges support from the Science and Technology Facilities Council (STFC; grant number ST/R000549/1) and start-up funds from the University of Leeds.
Astrophysics at Queen's University Belfast is supported by a grant from the STFC (ST/P000312/1). TJM thanks Leiden Observatory for hospitality.
\appendix
\begin{figure*}[htbp]
\begin{center}
\includegraphics[scale=0.6]{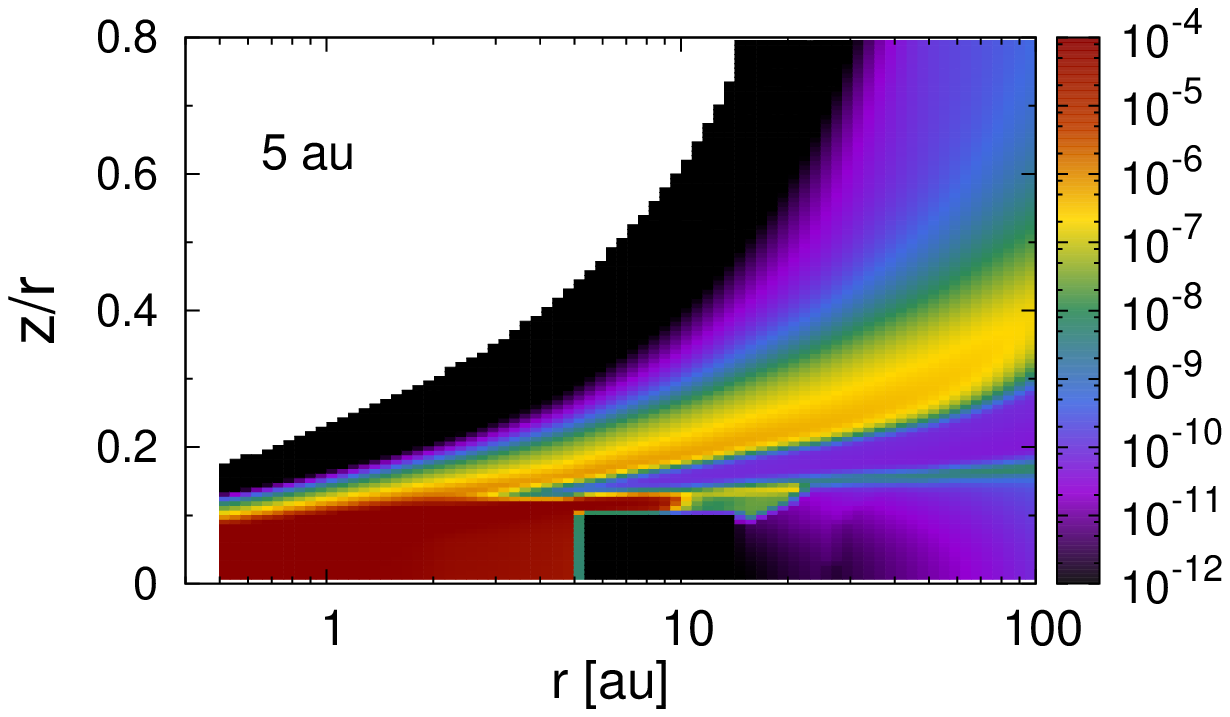}
\includegraphics[scale=0.6]{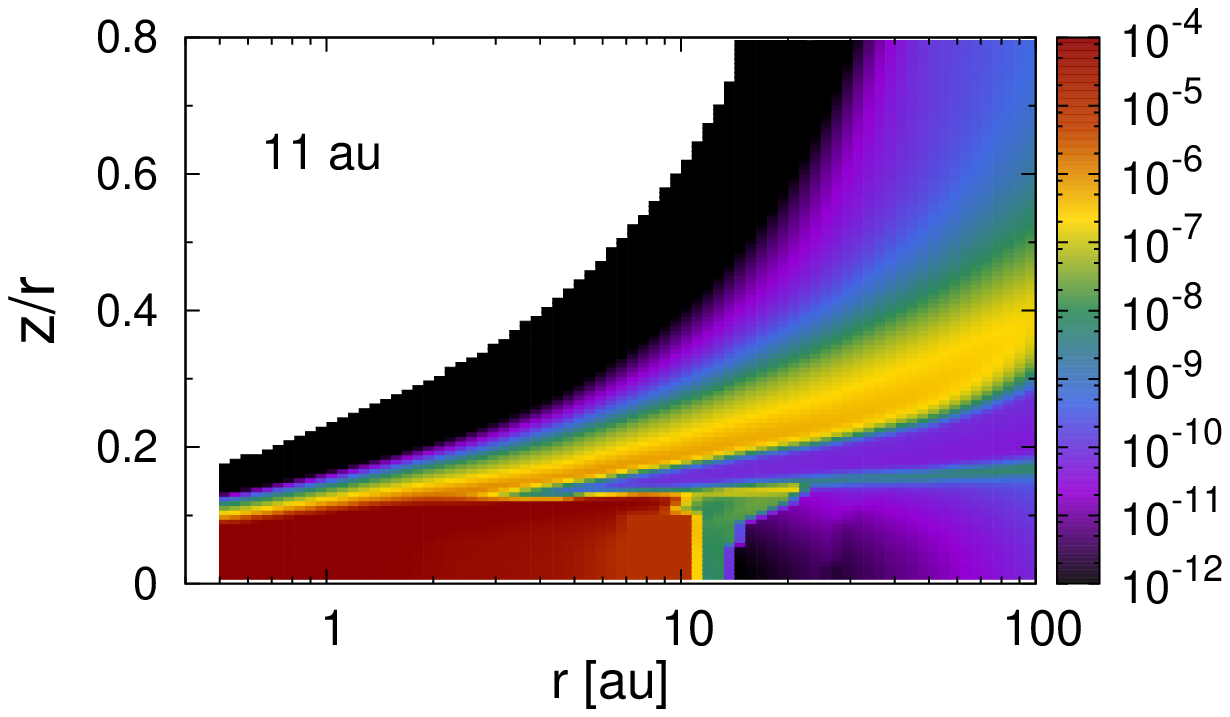}
\includegraphics[scale=0.6]{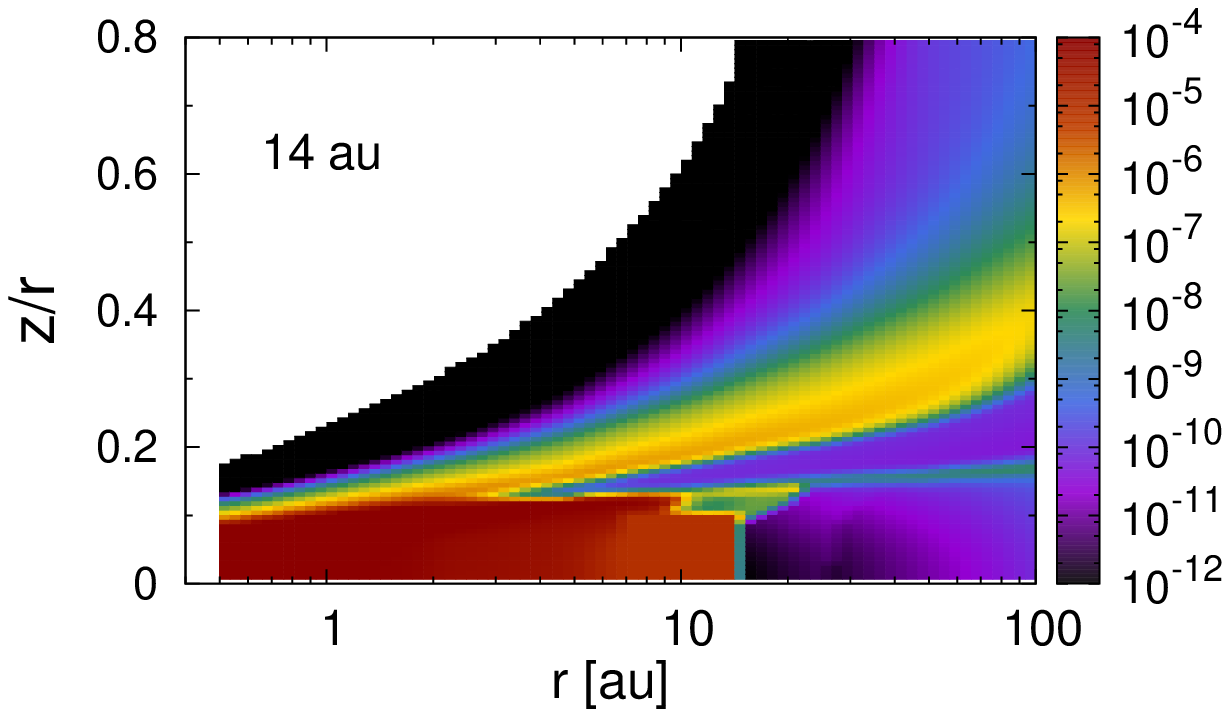}
\includegraphics[scale=0.6]{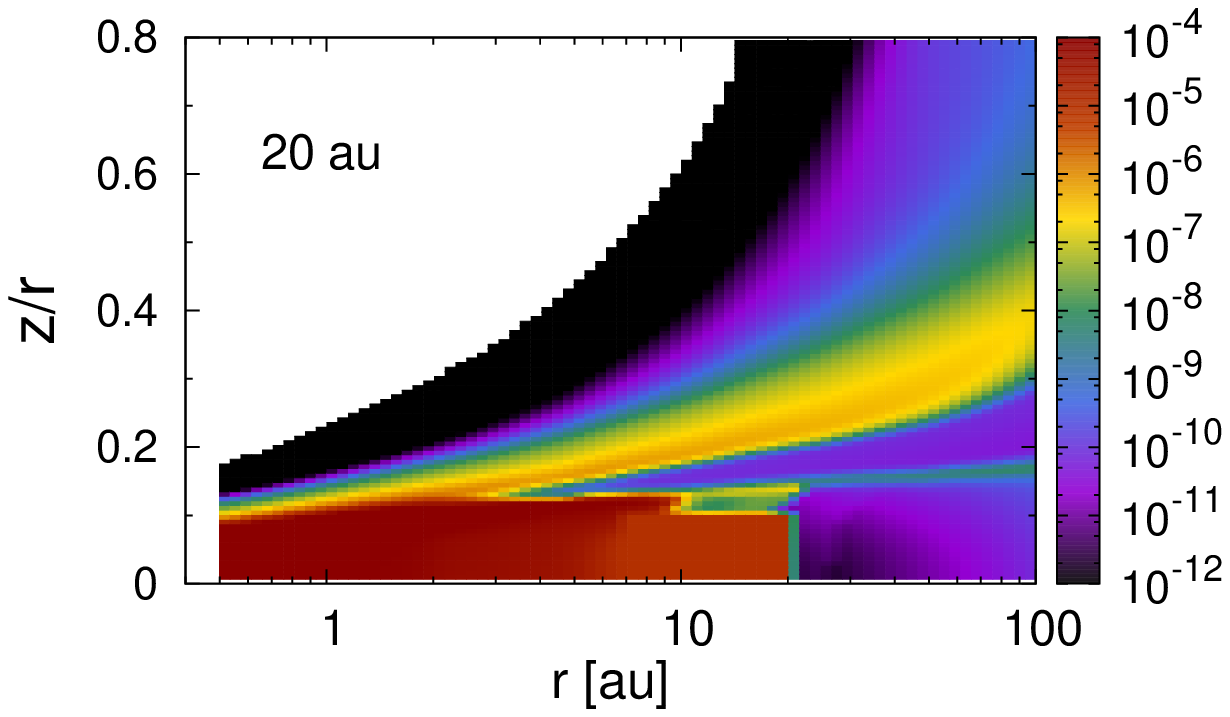}
\end{center}
\vspace{0.3cm}
\caption{\noindent 
The fractional abundance (relative to total hydrogen nuclei density) distributions of water gas of a disk around a Herbig Ae star as a function of the disk radius in au and height (scaled by the radius, $z/r$) up to a maximum radius of $r =100$ au.
In these plots, we fix the dust opacity and artificially change the outer edge of the region with a high $\mathrm{H_2}\mathrm{O}$ vapor abundance ($10^{-5}$) from 8 au (\citealt{Notsu2017, Notsu2018a}, see also Figure \ref{Figure4_add_paperIV} of this paper), to 5 au ($T_{g}\sim$180K, top left panel), 11 au ($T_{g}\sim$135K, top right panel), 14 au ($T_{g}\sim$120K, bottom left panel), and 20 au ($T_{g}\sim$100K, bottom right panel). 
\vspace{0.5cm}
}\label{Figure8_add_paperIV}
\end{figure*} 

%
%
\end{document}